\documentclass[preprint2,times,tighten]{aastex6}
\usepackage{amsmath,amstext}
\usepackage{url}
\usepackage{upgreek}
\usepackage[all]{hypcap} 

\shorttitle{A Versatile Technique to Achieve milli-Kelvin Stability for Precise RV}
\shortauthors{Stefansson et al.}
\newcommand{\unit}[1]{\ensuremath{\, \mathrm{#1}}}
\newcommand{\cms}{cm~s$^{-1}$}
\newcommand{\ms}{m~s$^{-1}$}

\begin{document}
\title{A Versatile Technique to enable sub-milli-Kelvin instrument Stability for Precise Radial Velocity Measurements: Tests with the Habitable-zone Planet Finder}
\author{
Gudmundur Stefansson\altaffilmark{1,2,3,4,5}, 
Frederick Hearty\altaffilmark{1}, 
Paul Robertson\altaffilmark{6,1,2}, 
Suvrath Mahadevan\altaffilmark{1,2,3}, 
Tyler Anderson\altaffilmark{1}, 
Eric Levi\altaffilmark{1}, 
Chad Bender\altaffilmark{1,2,7}, 
Matthew Nelson\altaffilmark{8}, 
Andrew Monson\altaffilmark{1}, 
Basil Blank\altaffilmark{9}, 
Samuel Halverson\altaffilmark{6,1,2,3,10}, 
Chuck Henderson\altaffilmark{11}, 
Lawrence Ramsey\altaffilmark{1,2}, 
Arpita Roy\altaffilmark{1,2,3}, 
Christian Schwab\altaffilmark{12}, 
Ryan Terrien\altaffilmark{1,2,3,13}
}
\email{gudmundur@psu.edu}
\thanks{Certain commercial equipment, instruments, or materials are identified in this paper in order to specify the experimental procedure adequately. Such identification is not intended to imply recommendation or endorsement by the National Institute of Standards and Technology, nor is it intended to imply that the materials or equipment identified are necessarily the best available for the purpose.}

\altaffiltext{1}{Department of Astronomy \& Astrophysics, The Pennsylvania State University, 525 Davey Lab, University Park, PA 16802, USA}
\altaffiltext{2}{Center for Exoplanets \& Habitable Worlds, University Park, PA 16802, USA}
\altaffiltext{3}{Penn State Astrobiology Research Center, University Park, PA 16802, USA}
\altaffiltext{4}{Leifur Eiriksson Foundation Fellow}
\altaffiltext{5}{NASA Earth and Space Science Fellow}
\altaffiltext{6}{NASA Sagan Fellow}
\altaffiltext{7}{Steward Observatory, University of Arizona, 933 N. Cherry Ave., Tucson, AZ 85719}
\altaffiltext{8}{Department of Astronomy, University of Virginia, 530 McCormick Rd, Charlottesville, VA 22904, USA}
\altaffiltext{9}{PulseRay Inc., Beaver Dams, NY, USA}
\altaffiltext{10}{Department of Physics and Astronomy, University of Pennsylvania, Philadelphia, PA 19104, USA}
\altaffiltext{11}{Cornell University, Ithaca, NY, USA}
\altaffiltext{12}{Macquarie University, Sydney, Australia}
\altaffiltext{13}{National Institute of Standards and Technology, Boulder, CO, USA}

\begin{abstract}
Insufficient instrument thermo-mechanical stability is one of the many roadblocks for achieving 10~\cms~ Doppler radial velocity (RV) precision, the precision needed to detect Earth-twins orbiting Solar-type stars. Highly temperature and pressure stabilized spectrographs allow us to better calibrate out instrumental drifts, thereby helping in distinguishing instrumental noise from astrophysical stellar signals. We present the design and performance of the Environmental Control System (ECS) for the Habitable-zone Planet Finder (HPF), a high-resolution (R=50,000) fiber-fed near infrared (NIR) spectrograph for the $10\unit{m}$ Hobby Eberly Telescope at McDonald Observatory. HPF will operate at $180 \unit{K}$, driven by the choice of an H2RG NIR detector array with a $1.7\unit{\upmu m}$ cutoff. This ECS has demonstrated $0.6 \unit{mK}$ RMS stability over 15 days at both $180 \unit{K}$ and $300 \unit{K}$, and maintained high quality vacuum ($<$$10^{-7} \unit{Torr}$) over months, during long-term stability tests conducted without a planned passive thermal enclosure surrounding the vacuum chamber. This control scheme is versatile and can be applied as a blueprint to stabilize future NIR and optical high precision Doppler instruments over a wide temperature range from $\sim$77$\unit{K}$ to elevated room temperatures. A similar ECS is being implemented to stabilize NEID, the NASA/NSF NN-EXPLORE spectrograph for the $3.5 \unit{m}$ WIYN telescope at Kitt Peak, operating at $300 \unit{K}$. A full SolidWorks 3D-CAD model\footnote{SolidWorks model: \url{https://scholarsphere.psu.edu/files/7p88cg66f}} and a comprehensive parts list of the HPF ECS are included with this manuscript to facilitate the adaptation of this versatile environmental control scheme in the broader astronomical community.

\end{abstract}
\maketitle

\section{Introduction}
One of the most exciting goals in exoplanet science is to detect and characterize an Earth-like planet in the habitable zone. The Doppler radial velocity (RV) amplitude of an Earth-like planet around a Sun-like star is $\sim$10~\cms, well below the currently achievable RV precision of modern high precision Doppler spectrometers.
Attaining these levels of precision will require an unprecedented understanding of both instrumental and astrophysical noise sources. The instrumental precision of these spectrometers is not set by a single error source, but rather by the contribution of many individual error sources, including illumination instabilities and calibration errors \citep{Halverson2016, Fischer2016}. Among these,  some of the instrumental noise sources that can be very design-intensive to solve are instrumental thermo-mechanical instabilities.
Temperature and pressure variations inside or around the spectrograph directly translate to mechanical warping in the optics and optical mounts and change of the dimensions and refractive indices of optical glasses, manifesting as changes in the illumination of the optics and resulting in variations in the instrument point spread function (PSF) \citep{Pepe2014Nature}. If not properly addressed, these PSF variations can easily result in Doppler errors over 100~\ms. Consequently, the RV exoplanet community is pushing towards building spectrographs that are intrinsically highly stable. These dedicated spectrometers will have the ability to precisely calibrate instrumental drifts, allowing for the separation of instrumental noise from astrophysical RV signals \citep{plavchan2015,Fischer2016}. In doing so, these intrinsically stable spectrometers will allow us to better decouple stellar activity from planet-signals (e.g., \cite{Robertson2014_g581,Robertson2014_g667c}), which remains one of the most significant challenges in precision Doppler velocimetry.

Enclosing the spectrometer optical train in a highly temperature stabilized vacuum chamber is a direct and effective way to minimize instrumental drifts.
High quality vacuum minimizes index of refraction variations of the air within the spectrograph, which would otherwise directly lead to pressure-induced RV errors. Furthermore, high quality vacuum eliminates thermal convective and conductive heat transport of energy from the gas within the spectrograph.
Temperature stabilization minimizes stresses between the optics and optical mounts caused by mismatches in thermal expansion coefficients (CTEs), reducing variations in the beam path.
Although temperature gradients will inevitably be present in systems with components operating at inherently different temperatures, proper temperature stabilization eliminates their variation with time, minimizing thermo-mechanical drifts.
The fiber-fed vacuum chamber approach for high Doppler precisions of $\sim$1~\ms~in the optical was pioneered by the HARPS spectrograph at La Silla Observatory in Chile. The HARPS vacuum chamber is temperature stabilized to $\pm 0.01 \unit{K}$ throughout the year and consistently maintains vacuum pressure below $\sim$$10^{-3} \unit{Torr}$ \citep{Mayor2003}. This extraordinary stability of HARPS is key to its success, allowing HARPS to routinely achieve RV precisions $\sim$1~\ms~\citep{Mayor2003}, and even down to the 0.7~\ms~level for carefully sampled stars \citep{Lovis2006}; HARPS has demonstrated that high levels of instrument stability translate directly to better RV precision \citep{Mayor2003}. Since the success of HARPS, stabilized spectrographs built for precision applications are now becoming standard practice, and spectrographs aiming at $1\unit{mK}$ temperature stability or better are now being built.

Achieving long-term precise environmental control has been a difficult problem---especially approaching the $1 \unit{mK}$ level---as many poorly understood, and hard-to-control short-and long-term effects start to dominate. Traditionally, spectrographs were attached to one of the telescope ports, subject to large day-to-night, day-to-day, and annual temperature swings. The introduction of optical fibers and double scramblers \citep{Hunter1992} to feed the starlight directly into the spectrograph, allowed the spectrograph to be stowed away in temperature stabilized rooms, away from the harsh telescope environment. Once it is no longer physically mounted to the telescope, a spectrograph's weight is no longer a driving issue, allowing for heavier construction materials with long thermal time constants to buffer out high frequency temperature variations. Recently, groups have moved towards designing fiber-fed diffraction limited spectrographs (e.g., \cite{Schwab2012,Crepp2014,Blake2015}) ---with smaller optics than the traditional seeing limited spectrograph---having great potential for increased thermal, pressure, and mechanical stability due to their overall smaller footprint. It is through heritage from past instruments, and new technology development, that we are now designing spectrographs to be stable at the $\leq$$1 \unit{mK}$ level.

Near-infrared (NIR) Doppler RV measurements have historically lagged behind optical measurements. This is due, in part, to numerous outstanding instrumental challenges associated with operating in the NIR, including lack of wavelength calibration sources, cryogenic operation, and the need to use NIR detector technology. Optical spectrographs can operate at room temperature, while NIR spectrographs have to be cooled to cryogenic temperatures to suppress thermal background radiation. To reach these cold temperatures, some high precision Doppler spectrometers use innovative continuous-flow gaseous nitrogen coolers (e.g., CARMENES \cite{Mirabet2014}), others LN2 tanks (e.g., APOGEE \cite{Wilson2010}), and others Stirling coolers (e.g., iLocater \cite{Crepp2014}). Furthermore, operation at cryogenic temperatures requires prolonged, high quality vacuum operation. This places constraints on the materials used in the spectrometer and subsystems, as materials with measurable outgassing properties must be excluded. NIR instruments are also harder to service and maintain as they can take a long time (from days to weeks) to reach operating temperatures. Cryogenic operation makes instrument alignment significantly more difficult, as typical materials used in spectrometer optical assemblies and mounts have non-zero CTEs. In spite of this added complexity, many NIR spectrometers are currently under development that aim for $\sim$1~\ms~precision, such as CARMENES \citep{Quirrenbach2012}, HPF \citep{Mahadevan2014}, iLocater \citep{Crepp2014}, IRD \citep{Kotani2014} and Spirou \citep{artigau2014}.

Here we present an environmental control technique to achieve milli-Kelvin thermal stability for precise radial velocity measurements. This environmental control technique is based on the Environmental Control System (ECS) for the Habitable-zone Planet Finder, a fiber-fed NIR spectrograph operating at $T=180 \unit{K}$, currently being built for the Hobby-Eberly Telescope (HET). This ECS has demonstrated $\sim$$0.6 \unit{mK}$ RMS thermal stability over 15 days on the optical bench, and $<$$10^{-7}\unit{Torr}$ pressure stability over months. Although the environmental control techniques discussed in this paper are optimized for HPF, this control scheme is versatile, and can be easily modified and used as a blueprint to stabilize astronomical instruments at a wide range of temperatures from $77\unit{K}$ to elevated room temperatures.

This paper is sectioned as follows.
Section 2 gives an overview of the HPF instrument.
Section 3 gives a brief overview of the top level design drivers for the HPF ECS. 
Further details are discussed in Section 4. 
Section 5 outlines the technical implementation of the various subsystems composing the HPF ECS. 
Section 6 describes the experimental setup for two long-term environmental stability tests of the HPF ECS operating at cryogenic, and elevated room temperatures, and the main results from these tests are presented in Section 7.
Section 8 discusses the versatility of this ECS for operation at a wide range of temperatures, and presents final conclusions.


\section{The Habitable-zone Planet Finder}
The Habitable-zone Planet Finder is a stabilized, fiber-fed, cross-dispersed high-resolution ($\textrm{R} = 50,000$) NIR spectrograph, a facility-class instrument for the $10 \unit{m}$ Hobby-Eberly Telescope (HET) at McDonald Observatory \citep{Mahadevan2012,Mahadevan2014}. The optical design of HPF is based on an asymmetric white pupil relay covering the information rich z, Y and J NIR bands ($800-1300 \unit{nm}$). The spectrograph optics are kept inside a vacuum cryostat cooled to $180 \unit{K}$, motivated by the choice of a Hawaii-2RG (H2RG) NIR detector from Teledyne with a $1.7 \unit{\upmu m}$ cutoff. HPF uses the simultaneous reference technique, and has three fibers that record science, calibration and sky spectra simultaneously. HPF draws rich design heritage from the Penn State Pathfinder NIR prototype \citep{Ramsey2010,Ycas2012}, the APOGEE instrument \citep{Wilson2010,Blank2010}, and the PRL Stabilized High Resolution Echelle fiber-fed Spectrograph (PARAS) \citep{Chakraborty2014}, and is in turn a significant source of heritage for the NN-EXPLORE NEID spectrometer being designed for the $3.5 \unit{m}$ WIYN telescope\footnote{\url{http://neid.psu.edu/}}. HPF has a goal RV precision of 1~\ms, with a 3~\ms~requirement, in order to meet its science goal of surveying nearby M-dwarfs for rocky planets in the habitable-zone of mid-to-late M-dwarfs. The on-sky HPF survey plan is described in \cite{Mahadevan2012}.

\begin{figure}[t]
	\begin{center}
		\includegraphics[width=\columnwidth]{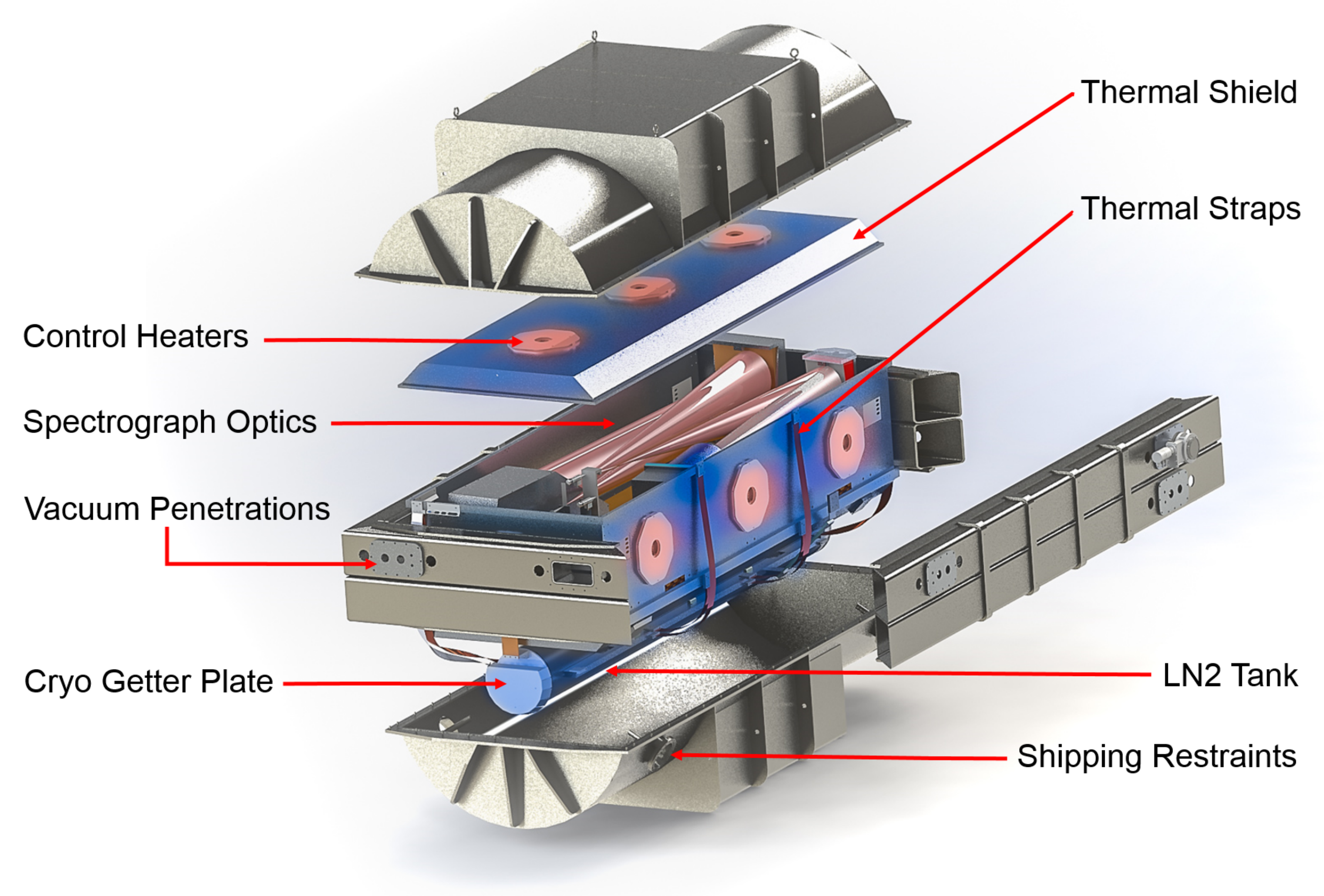}
	\end{center}
	\caption{An overview diagram of the HPF cryostat. The cryostat is responsible for keeping the optics under high quality vacuum, necessary for long-term stable operation at the $180 \unit{K}$ operating temperature. The cryostat supplies the necessary vacuum penetrations for optical fibers, electronics cables, and supplies low-conductivity mounting points for the optical bench and LN2 tank. An actively temperature controlled radiation shield surrounds the optical train.}
	\label{fig:hpfcryostat}
\end{figure}

A solid model diagram of the HPF vacuum cryostat is shown in Figure \ref{fig:hpfcryostat}. The cryostat is derived from the APOGEE cryostat design \citep{Wilson2010,Blank2010,Hearty2014}, with key modifications to achieve high temperature stability. The entire optical train and thermal shield are kept under high quality vacuum ($<$$10^{-7}\unit{Torr}$) to support prolonged operation and stable operation at the $180 \unit{K}$ operating temperature \citep{Hearty2014}. HPF will be housed in a dedicated thermal enclosure in the HET Spectrograph Room, which provides a passive thermal buffer to suppress high frequency thermal fluctuations from the observatory HVAC system.

Since its delivery to the Penn State Integration Lab in Summer 2015, the HPF cryostat and ECS systems have been in integration-and long-term-verification phase. Environmental stability tests were performed both at the cold $180 \unit{K}$ operating temperature of HPF, and also at an elevated room-temperature of nominally $300 \unit{K}$. The latter test used HPF as a prototype test case for the NEID spectrograph to empirically demonstrate the performance of this ECS at warmer temperatures. 

\section{Environmental Stability Design Drivers}
High precision RV spectrographs have stringent pressure, temperature, and mechanical stability requirements. For fiber-fed spectrographs, some of these effects can be traced and removed using a simultaneous calibration fiber. However, in practice it is best to keep the instrument as absolutely stable as possible. These aspects are discussed in general in the following subsections. The next section includes details driving the HPF ECS design. 

\subsection{Pressure stability}
Pressure changes of the air within the spectrograph induce refractive index variations, varying the wavelength of the light propagating through the spectrograph, as dictated in Equation 1:
\begin{equation}
\lambda = \frac{\lambda_{\textrm{vac}}}{n(P,T)},
\label{eq:lambda}
\end{equation}
where $\lambda_{\textrm{vac}}$ is the wavelength in vacuum, and the refractive index $n(P,T)$ as a function of pressure and temperature is given by the Edlen equation \citep{edlen1966}. Varying wavelengths correspond to RV errors, given by the Doppler equation,
\begin{equation}
\Delta \textrm{v} = \frac{\Delta \lambda }{ \lambda_{\textrm{vac}}} c.
\label{eq:rvdoppler}
\end{equation}

Using Equations \ref{eq:lambda} \& \ref{eq:rvdoppler}, we calculate that a typical day-to-day atmospheric fluctuation in pressure of $3 \unit{Torr}$ corresponds to an RV error of $\sim$300~\ms~at $T = 180 \unit{K}$. Similarly, we calculate that a variation in pressure of $\Delta P = 10^{-6}\unit{Torr}$, corresponds to an RV error of $\sim$0.02~\cms, where we see that pressure variation effects become a negligible factor in the RV error budget; these are the vacuum pressure levels that can be reached with current and upcoming spectrographs. We stress that although these vacuum levels may exceed requirements for measurements at these precisions, this level of absolute vacuum pressure is important for maintaining thermal stability, as discussed below.

For spectrographs with subsystems operating at inherently different temperatures---e.g., spectrographs employing heaters and cold thermal sinks---absolute vacuum pressure of the gas within the vacuum chamber represents one of the key hurdles for achieving thermal stability at the milli-Kelvin level. This is due to molecular conduction of energy: the gas within the vacuum chamber thermally short-circuits the otherwise thermally isolated components of the spectrograph. This is a different situation from the isothermal approach of the HARPS spectrograph, which keeps the whole optical train under moderate vacuum ($\sim$$10^{-3} \unit{Torr}$) and at the same temperature as the vacuum chamber. Long-term temperature stability tests using a scale-model version of the HPF cryostat described in \cite{Hearty2014} show that increasing the vacuum pressure above $\sim$$10^{-5} \unit{Torr}$ level resulted in temperature drifts substantially larger than at pressures $<$$10^{-6} \unit{Torr}$. For this reason, the HPF ECS is designed to operate at an absolute, long term vacuum pressure of $<$$10^{-6} \unit{Torr}$, eliminating this additional source of temperature variability.

\subsection{Temperature stability}
Temperature fluctuations within the spectrometer environment can be a major source of RV error if not properly minimized. There are many factors that drive the achievable instrument stability. The primary goal is to minimize differential thermal contractions and expansions between the optical mounts and optics themselves, which induce wavelength dependent variations in the beam path, manifesting as spectral shifts on the detector.

One adverse side effect of imprecise thermal control is the change in groove density in the echelle grating due to thermal fluctuations. This change results in a measurable spectral shift in the focal plane, potentially mimicking an astrophysical Doppler shift. For a uniform change in temperature $\Delta T$ on the grating, this RV error can be shown to equal,
\begin{equation}
\Delta v = \alpha_L c \Delta T,
\label{eq:grating}
\end{equation}
where $\alpha_L$ is the CTE of the grating material. Therefore, echelle gratings are usually fabricated out of low CTE materials such as Schott Zerodur, which minimizes thermal sensitivity of the diffraction grooves. At room temperature Zerodur has $\alpha_L \leq 10^{-7} \unit{K}$, leading to an approximate RV error of $\leq$$3 \unit{cm~s^{-1}mK^{-1}}$ based on Equation \ref{eq:grating}. The CTE of Zerodur has also been extensively studied as a function of temperature. At the $180 \unit{K}$ operating temperature of HPF, the class of Zerodur used in the HPF grating substrate has an $\alpha_L \sim 2 \cdot 10^{-7} \unit{K}$, leading to a thermal sensitivity of $\sim$$6 \unit{cm~s^{-1}mK^{-1}}$. 

Additionally, the groove density of the grating can also change if the echelle substrate material varies in dimension over long timescales. This is a known effect with Zerodur, and has been documented in the literature by \cite{Bayer-Helms1985}, who looked at the dimensional stability of Zerodur over multiple decades. They characterize the aging of Zerodur with an aging coefficient $A$, 
\begin{equation}
\frac{\Delta l}{l} = A \Delta t,
\label{eq:aging}
\end{equation}
where $\Delta l /l $ is the relative shrinkage during the time interval $\Delta t$. The value of $A$ depends on the annealing rate of the Zerodur slab, and monotonically decreases with its age. In their sample, $A$ has a value between $-0.69 \times 10^{-6} \unit{year^{-1}}$ and  $-0.03 \times 10^{-6} \unit{year^{-1}}$ \citep{Bayer-Helms1985}. The aging coefficient of Zerodur can then be related to an RV change by,
\begin{equation}
\Delta v = c A \Delta t,
\label{eq:rvaging}
\end{equation}
corresponding to a $9 \unit{m~s^{-1} year^{-1}}$ to $200 \unit{m~s^{-1} year^{-1}}$ measurement drift due to this shrinkage effect of Zerodur. 

Although the two effects discussed above both alter the echelle grating grove density, we emphasize that they happen on very different timescales. If proper thermal control is lacking, the echelle groove density variations due CTE effects happen rapidly, and can thus affect the calibration and science light in differing ways, leading to a less effective calibration. The latter aging effect is a slow and monotonically varying process, and thus more likely to be easily traced by daily calibrations and corrected for in the RVs.

The calculations above have focused on the echelle grating. The total RV shift observed for a given temperature change in the total spectrograph system is a complex problem, and often difficult to model precisely. To gain further insight into this complex behavior, the spectrograph layout can be simulated---at least partially---using 3D Finite Element Analysis (FEA) modeling software such as SolidWorks, to calculate the contraction/expansion effects on the optics and optical mounts for a given temperature offset. The beam path can then be traced through the perturbed optical train and the wavelength shifts on the detector calculated (e.g., through the generation of a synthetic spectrum such as described in e.g., \cite{Gibson2016}). These results can then be folded into a comprehensive error budget overview, such as described by \cite{Podgorski2014}, and \cite{Halverson2016}, to evaluate the impact of the thermal change on the overall RV error budget. In practice, it is difficult to simulate a fully realistic model of the spectrograph layout, due to uncertainties in the exact mechanical and material properties of the as-built spectrograph, transient phenomena, and the large computing resources required to model such a large system at high resolution. However, even a simple model---specially if tied to a prototype system or a concept demonstration---can be extremely useful, especially during the design phase, to get an estimate of the RV stability of the spectrograph with temperature.

\section{HPF Environmental Control Design Drivers}
The exact requirements on temperature and pressure stability flow down from the 1~\ms~goal RV precision of HPF. The requirements were calculated using the equations described above, alongside careful trade-studies.

The choice of the H2RG NIR detector with a $1.7 \unit{\upmu m}$ cutoff drives the choice for the $180 \unit{K}$ operating temperature of the HPF radiation shield and optical train. This operating temperature---approximately halfway between LN2 temperature and ambient room temperature---was chosen as it sufficiently minimizes background blackbody radiation for the $1.7 \unit{\upmu m}$ cutoff of the H2RG detector, without making the optical train fully cryogenic. The detector itself is designed to operate at low noise at $\sim$$120 \unit{K}$ and is directly coupled to the LN2 tank with an aluminum cold-finger with a dedicated copper thermal strap.

From an instrument perspective there is a strong desire to operate at temperatures close to room temperature as opposed to at cryogenic temperatures. This is due to many factors. First, the absolute value of the CTE of the Zerodur echelle grating (assuming standard class Zerodur) is smaller at temperatures close to room temperatures\footnote{Schott Zerodur CTE: \url{http://www.schott.com/advanced_optics/english/download/schott_zerodur_katalog_july_2011_en.pdf}}, minimizing the impact the echelle has on the RV precision, as described by Equation \ref{eq:grating}. Second, the overall optical design, including the choice of glass and design of optical mounts, along with optical alignment is simpler if the final operating temperature is closer to assembly temperature. Last, cool-down and warm-up timescales are significantly shorter, reducing overall technical and schedule risks as testing cycles are quicker to perform.

In contrast to APOGEE, HPF has active control elements within the vacuum boundary distributed across the thermal shield, producing a near blackbody cavity for the full optical train. The temperature control electronics were initially developed and tested using a small-scale vacuum cryostat at The University of Virginia to verify low-noise and stable operation long-term \citep{Hearty2014}. Subsequently, this system has been further improved for ease of maintainability, and robust operation at the observatory. Section \ref{sec:technical} contains the technical implementation of the overall ECS, as well as descriptions of its subsystems.

\begin{figure*}[t]
	\begin{center}
		\includegraphics[width=\textwidth]{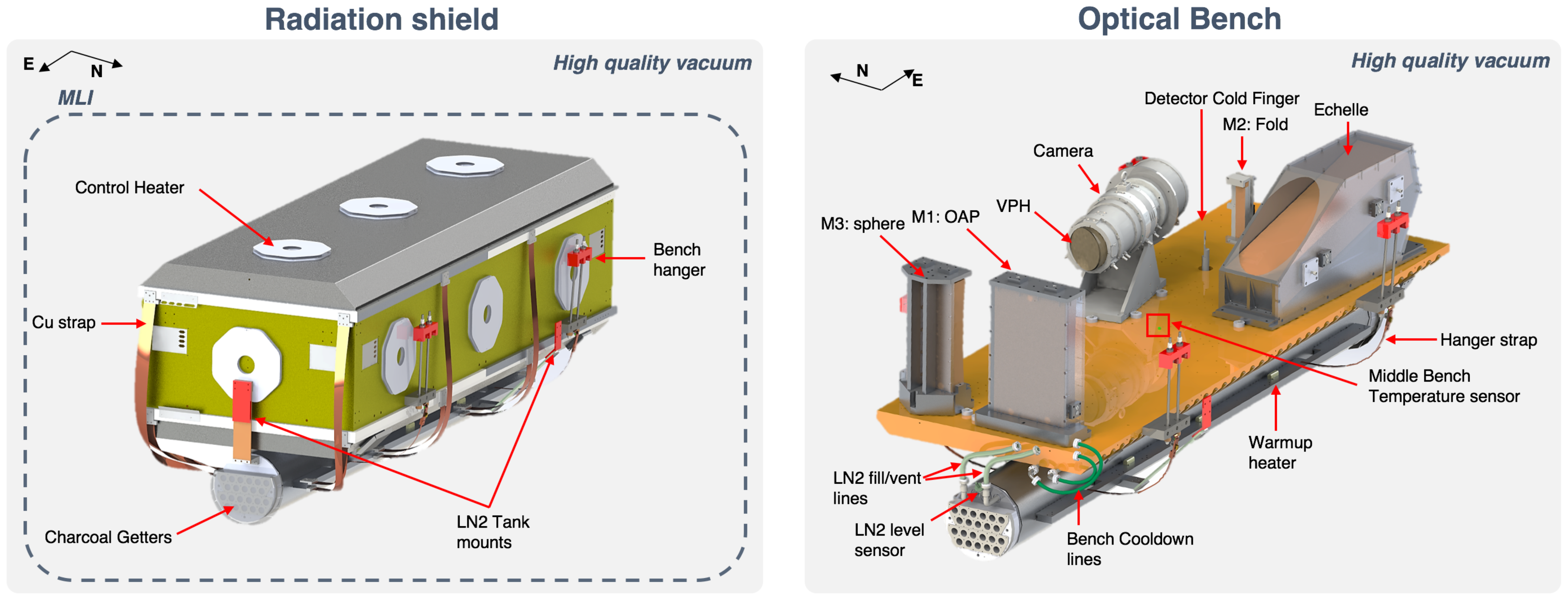}
	\end{center}
	\caption{Overview of the HPF ECS, and various subsystems that operate within the cryostat vacuum boundary. All of the mounting points to the cryostat (radiation shield hanger mounts, and LN2 tank mounts) are shown in red, and are insulated with G10. \textbf{Left:} An overview of the HPF ECS control boundary, showing the radiation shield, the locations of the surface mounted control heaters, LN2 tank, copper thermal straps, and the location of charcoal getters. The radiation shield, LN2 tank, and thermal straps all have individual MLI blankets. \textbf{Right:} The optical bench, along with other ECS subsystems and spectrograph optics. The temperature sensor on the middle of the bench is highlighted.}
	\label{fig:ecs_overview}
\end{figure*}

\section{Technical Implementation of the HPF ECS}
\label{sec:technical}

\subsection{Overview}
Figure \ref{fig:ecs_overview} shows a detailed view of the overall environmental control concept. The environmental control approach for HPF revolves around stabilizing the optics by enclosing them in a thermally stabilized radiation shield kept under high quality vacuum, as shown in Figure \ref{fig:hpfcryostat}. The high quality vacuum ($<$$10^{-7} \unit{Torr}$) removes both molecular convective and conductive transport of air within the spectrograph. Fundamentally, the radiation shield is used to keep the optical train within a fixed-temperature black-body cavity, by actively controlling the temperature of the radiation shield walls, while insulating it from ambient fluctuations with Multi-Layer Insulation (MLI) blankets \citep{Hearty2014}.

The optics sit on a massive flat aluminum 6061-T6 optical bench with long thermal time constants and high thermal conductivity to minimize thermal gradients and transient instabilities. The bench has mounting points for a set of 4 carefully designed hanger assemblies, composed of stainless steel and thermally isolated using G10, which mount the bench to the cryostat in a non-rigid way with low thermal conductivity. A set of 16 copper thermal straps couple the radiation shield to the LN2 tank, sized to over-cool the radiation shield down to nominally $170 \unit{K}$, approximately $10 \unit{K}$ below the final operating temperature. An additional set of carefully sized thermal straps, coupled directly to the LN2 tank, are attached to the optical bench hanger assemblies to compensate for the positive flow of conductive heat. The control heaters are responsible for keeping the radiation shield stable at the $180 \unit{K}$ operating temperature long term. They can be carefully tuned for any given temperature setpoint with milli-Kelvin resolution using the temperature control electronics. The optical bench and LN2 tank are outfitted with a dedicated warm-up and cool-down system to reduce the duration of thermal cycles, without cooling/warming up the optics too fast.

The ECS subsystems are further described below. To supplement the discussion in the following subsections, a full SolidWorks 3D-CAD model of the HPF ECS is available, hosted at the Penn State Scholarsphere data hosting service\footnote{\url{https://scholarsphere.psu.edu/files/7p88cg66f}}. Additionally, a comprehensive parts list of the HPF ECS is included in the Appendix (Table \ref{tab:partslist}1).

\subsection{A Robust Vacuum Cryostat}
\label{sec:cryostat}
The HPF cryostat (Figure \ref{fig:hpfcryostat}) is modeled after the APOGEE cryostat, proven to be a long-term stable and highly functional system \citep{Blank2010,Wilson2010}. The cryostat was modified in several ways to meet HPF's enhanced mechanical and thermal stability performance specifications. Support systems, such as vacuum support components, LN2 level indication and auto-fill systems, and cryostat mechanical supports are near identical. The system was fabricated by PulseRay (Beaver Dams, NY) and Cameron Machining (NY). 

The cryostat shell is fabricated out of $\unit{304}$ stainless steel (Figure \ref{fig:hpfcryostat}). This material is easy to weld and structurally strong, giving low deformations and stress points in finite element analysis models (see \cite{Hearty2014}). The overall dimensions of the cryostat are 3.1m long by 1.4m wide by 1.4m high, with a gross weight including optics about $2800 \unit{kg}$, providing for long thermal time constants. The cryostat has removable cylindrical lids on the top and bottom to ease maintainability, and to facilitate access to the optics, and mechanical and electrical systems. Commercially available box beams serve as the central frame of the vacuum chamber, providing surfaces for sealing (Viton O-rings) and mounting the upper and lower hoods, vacuum penetrations, and surfaces for supporting the instrument on air legs, rigging points etc. During operations the middle frame stands on 4 vibration isolation legs from Newport to suppress mechanical vibration in the HET spectrograph room, providing insurance against unanticipated instrument sensitivity to vibration. The cryostat has specifically engineered restraints for the optical bench for use during shipping: a set of four steel shafts with Trantorque bushings are used to lock the optical bench in place. Additionally, a specifically designed vibration damped stainless-steel shipping pallet with a set of four high-pressure air-skids is used during transport of the instrument.

%
\subsection{Radiation Shield}
\label{sec:radshield}
The main purpose of the radiation shield (Figure \ref{fig:ecs_overview}) is to provide the optics and optical bench with a stable radiative environment. The radiation shield is suspended from the optical bench using 24 low thermal conductivity stainless steel shoulder bolts, distributed along the lower radiation shield side-walls. 12-layer MLI blankets are used to minimize radiative coupling between the radiation shield and the ambient surroundings. The stable radiative environment is achieved by actively compensating for ambient temperature fluctuations using 14 surface mounted control heaters, each of which responsible for independently controlling a section of the radiation shield. The radiation shield is cooled by the copper straps to nominally $170 \unit{K}$. From there, the radiation shield is actively heated up $10 \unit{K}$ by the surface-mounted resistive control heaters to the final $180 \unit{K}$ operation temperature. At thermal equilibrium, almost no conductive heat flow is experienced between the bench and the shield, due to the minimal temperature difference between the two. 

The radiation shield is designed to minimize thermal gradients across its surface, so the optics see a homogeneous temperature cavity \citep{Hearty2014}. For this purpose, the radiation shield is fabricated out of 3003 aluminum, an alloy well known for its high thermal conductivity properties over a wide range of temperatures. The radiation shield walls are 1/8" ($\sim$$3 \unit{mm}$) thick. The attachment points for the Cu thermal straps are thick rails, to evenly distribute the heat-sink across the radiation shield. Additionally, a careful study presented in \cite{Hearty2014} of the exact location and number of control heaters and thermal straps was performed to minimize gradients, while keeping the number of control areas manageable. In the final configuration the modeled thermal gradients observed across the radiation shield are $0.6 \unit{K}$ from the coldest to the hottest spot. Although this level of thermal gradients is observed, the key is to keep them as static as possible, to provide the optical system with an unchanging radiative environment, eliminating thermal drivers that could change the temperature of the system. 

\subsection{Optical Bench and Optical Mounts}
\label{sec:opticalbench}
To ensure uniform thermal and mechanical performance, the optical bench was fabricated from a single, monolithic piece of 6061-T6 aluminum. The optical bench is light-weighted to minimize mass, yet avoid sagging, allowing for an overall RMS flatness of $<$$25 \unit{\upmu m}$ across the whole surface. The optical bench is suspended from the cryostat middle frame using a set of stainless steel hanger rods as is shown in Figure \ref{fig:ecs_overview}. The hanger assembly is insulated with G10 low thermal conductivity fiber-glass epoxy, minimizing the conductive heat input at the $\sim$$0.5 \unit{W}$ level per hanger. Dedicated copper straps (see Figure \ref{fig:ecs_overview}, right) are sized to short-circuit this heat input to the LN2 tank. The flat optical bench is machined with dedicated holes for optical mounts, making for straight-forward integration of optics.

The optical mounts (see Figure \ref{fig:ecs_overview}, right) are fabricated out of the same aluminum material as the bench (6061-T6) to mitigate differential flextures due to CTE mismatch effects. The mounts are aligned warm, then deliberately offset to compensate for CTE-induced contractions as the instrument is cooled down to the $180 \unit{K}$ operating temperature. The exact operating setpoint temperature can be tuned at the milli-Kelvin level to tune focus on the detector, allowing us to optimize image quality. The bench hangers have smooth half-dome shaped knuckles to accommodate the contraction of the bench, without inducing additional stresses when operating cold.

\subsection{Thermal Straps}
\label{sec:thermalstraps}
Copper thermal straps are used to cool down various components of the HPF ECS, by coupling the subsystems to the LN2 tank heat sink. A set of 16 thermal straps (see Figure \ref{fig:ecs_overview}) are sized to over-cool the radiation shield down to nominally $170 \unit{K}$. Additionally, each hanger mount assembly, which connects the optical bench to the cryostat, has one copper thermal strap each, sized to cancel out the conductive heat input to the bench. Additionally, a copper strap is used to couple the LN2 tank cold finger (see right panel in Figure \ref{fig:ecs_overview}) to the H2RG detector assembly to cool the detector down further to support operation at $120 \unit{K}$.

\begin{figure}
	\begin{center}		\includegraphics[width=\columnwidth]{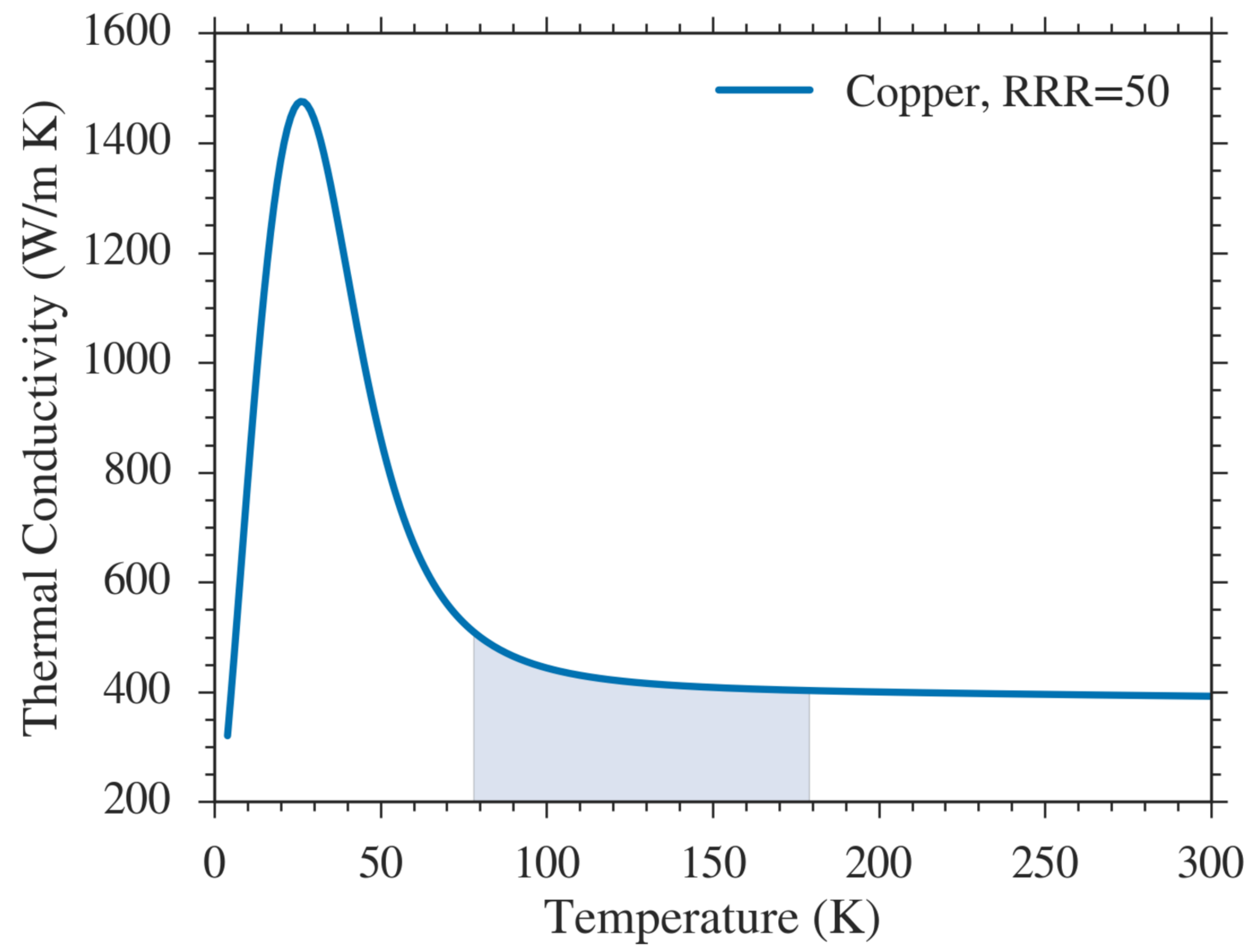}
	\end{center}
    \vspace{-0.5cm}
    \caption{Thermal conductivity of copper, $k(T)$, as a function of temperature, assuming a Residual Resistance Ratio of $\textrm{RRR=50}$. During operations the HPF copper straps will experience a temperature gradient from $\sim$$77 \unit{K}$ to $\sim$$180 \unit{K}$ (blue shaded area), with a non-negligible change in thermal conductivity. Data are obtained from the NIST cryogenics website; figure adapted from \cite{Mahadevan2014}.}
	\label{fig:cuconductivity}
\end{figure}

The thermal conductivity of copper, $k(T)$, varies non-negligibly with temperature (Figure \ref{fig:cuconductivity}). Therefore, the copper straps are sized for a heat draw $H_{\mathrm{Cu}}$:
\begin{align}
	H_{\mathrm{Cu}} = \frac{A}{L} \int_{T_C}^{T_H} k(T) dT,
	\label{eq:cucryo}
\end{align}
where $L$ and $A$ are the length and cross-sectional area of the thermal strap, respectively, $T_C \sim 77 \unit{K}$ is the fixed temperature of the LN2 tank, and the $T_H$ is the equilibrium temperature of the component to be cooled. For the radiation shield, the copper straps are sized to give a $T_H \sim 170 \unit{K}$, when the heat draw $H_{\mathrm{Cu}}$ matches the heat input from ambient radiation to the radiation shield, calculated using the Stefan-Boltzmann law. This input heat is largely governed by the quality of the MLI blanketing, discussed below.

\subsection{LN2 Tank}
\label{sec:ln2tank}
The LN2 tank (see Figure \ref{fig:ecs_overview}) acts as the ultimate heat-sink. The LN2 tank is fabricated out of 6061-T6 aluminum alloy, cylindrical in shape with welded end caps, and two long side-rails that act as attachment points for thermal strapping and the warm-up heater system. The tank is about $\sim$$110 \unit{L}$ in volume, with an LN2 hold time of 4-5 days. The tank is suspended from the cryostat middle frame using G10 low thermal conductivity fiber-glass epoxy.

Ambient pressure fluctuations cause fluctuations in the saturation temperature of LN2: a $1 \unit{Torr}$ change in ambient pressure corresponds to $100 \unit{mK}$ in the saturation temperature. Such deviations in ambient pressure happen frequently at the observatory on hour-timescales. Without addressing this issue, this directly relates to a change of the temperature of the radiation shield, causing the optics and optical bench to drift in temperature. Therefore, to avoid ambient barometric pressure effects, the HPF ECS employs a back-pressure regulator system to stabilize the LN2 saturation temperature. The LN2 boil-off gas is passed into a manifold to warm the cold gas, and then through an absolute pressure-referenced back-pressure regulator, maintaining the pressure stable at $780 \unit{Torr}$, a pressure higher than the maximum expected atmospheric variability at the observatory. 

\subsection{Warm-up and Cool-down Systems}
\label{sec:warmup}
The HPF ECS has a built-in warm-up heater system, and a forced cool-down system. These systems are exclusively used during thermal cycling, and their purpose is to reduce instrument maintenance times, minimizing impact on the science survey, in case the instrument needs to be serviced.

The purpose of the warm-up heater system is to warm up the optical bench and LN2 tank from cold operating temperatures in a practical time frame of $\sim$$24$ hours during testing and maintenance. Additionally, the heaters on the optical bench allow for tuning of the temperature of the optical bench, to quickly reach the operating temperature. The warm-up heater system is composed of 16 Vishay Dale $50 \unit{\Omega}$ resistive heaters with a $50 \unit{W}$ power-rating, mounted in symmetrical pairs on the optical bench (5 pairs), and on the side-rails of the LN2 tank (3 pairs). Each pair of resistors is wired in parallel to give an effective resistance of $25 \unit{\Omega}$. Two manually controlled Variac transformers supply a tunable voltage to each heater pair. This system is kept completely separate from the active heater control system, and disabled during normal operation of the instrument to prevent accidental instrument heating.

Similarly to the warm-up heater system, the HPF ECS has a dedicated forced cool-down system to cool down the massive optical bench at a few degrees per hour, at a rate slow enough to not risk the integrity of the optics, while still allowing the bench to cool down within a reasonable time-frame. Without this system, the cool-down rate of the bench would be impractically long. The forced cool-down system consists of a copper line (see right panel in Figure \ref{fig:ecs_overview}) heat-sunk to the under side of the optical bench, which is then fed by LN2 during cooling. Once a forced cool-down is completed and the bench is near the $180 \unit{K}$ operating temperature, the gas in the copper lines are pumped out, to minimize conductive heat flow between the bench and the hermetics on the cryostat shell. In practice, we observe that the bench cool-down lines provide a $\sim$$1 \unit{W}$ conductive heat input to the bench. This is corrected for by sizing the hanger straps accordingly to match this heat input.

\subsection{Thermal Insulation}
\subsubsection{Multi Layer Insulation Blankets}
\label{sec:mli}
The ECS uses Multi-Layer Insulation (MLI) blankets to thermally insulate critical components such as the radiation shield, liquid nitrogen tank, copper straps, and detector cold finger from the surrounding environment. MLI blankets primarily reduce heat loss by thermal radiation, making them efficient thermal control elements for cryo-cooled instruments kept under high-vacuum. The HPF MLI blankets are similar in design to the blankets used for the APOGEE spectrograph, composed of multiple alternating thin sheets of aluminized Mylar (highly reflective), and a netted spacer material (nylon Tulle). Most of the HPF blankets have 12 Mylar layers, providing an effective emissivity of $\sim$$0.007$.

\begin{figure}
	\begin{center}
		\includegraphics[width=\columnwidth]{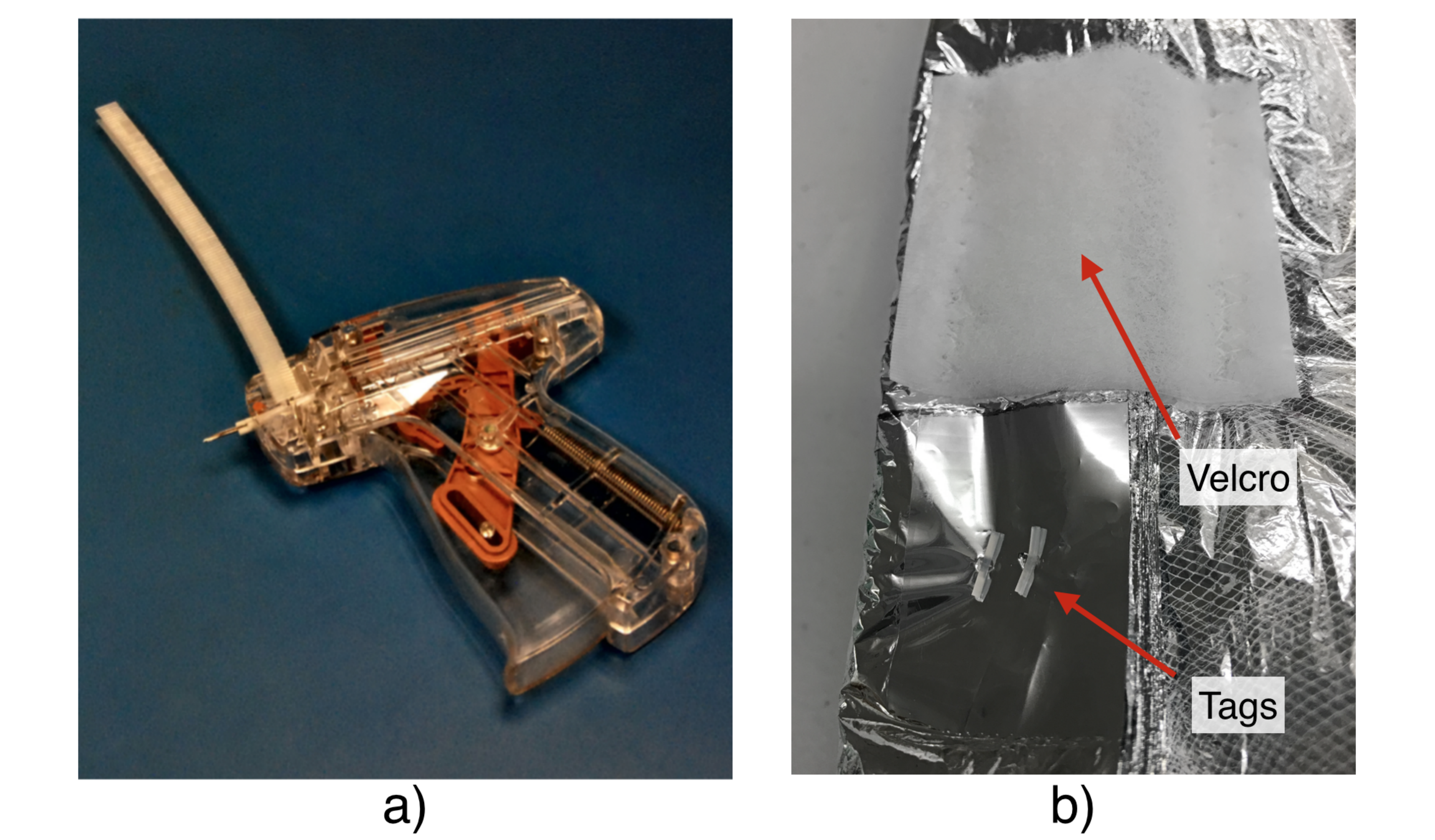}
	\end{center}
	\caption{a) Tag-gun with nylon I-pins used in the fabrication of the HPF MLI blankets; b) A photo of two I-pins, tagged through a patch of aluminized tape on one of the HPF MLI blankets for added strength. Also shown is a sewn-on Velcro patch. Left image adapted from \cite{Mahadevan2014}.}
	\label{fig:mli_tag}
\end{figure}

Traditionally, MLI blankets are sewn together, and the multi-layered blanket held together by stitches. However, any kind of hole or puncture in the blanket degrades its overall thermal performance. Another method, uses small tag-pins---small nylon "I" shaped tags used to hook price tags to clothes---to fix the layers in place (see e.g., \cite{hatakenaka2013}). This greatly reduces the amount of holes in the blankets, and tagging a few inches between tags is faster and less error-prone than sewing around the whole perimeter of the blanket. Moreover, the tag-pins allow one to fasten the layers together without compressing them, which reduces stress around the holes \citep{hatakenaka2013}. Lastly, the blankets tend to contract in the direction of sewing, which might lead them to be too small if not accounted for properly. All of the HPF blankets used this tag-pin method. Figure \ref{fig:mli_tag} shows the tag-gun used for the HPF blankets, along with one of the tags holding together a blanket: aluminized tape was used for every tagged area for added strength, as the thin aluminized Mylar tears easily. A video further explaining the HPF MLI fabrication process can be found online\footnote{\url{http://hpf.psu.edu/2014/09/29/mli-blankets/}}.

The HPF blankets are sized, aligned, and held together to cover the whole radiation shield, liquid nitrogen tank, and copper thermal straps. Strategically placed Velcro-pads are used to align and hold the blankets in place on the instrument. The Velcro pads are sewn on the blankets to strongly fix them to the blankets. This results in more holes punched on the area of the Velcro than the tagging, but sewing fixes them securely to the blankets.

All of the blankets are folded at areas where two blankets overlap: if two MLI blankets touch, care is taken to only make the outermost Mylar layer touch a corresponding outermost layer of the other blanket (or corresponding inside-to-inside). Touching outside-to-inside layers thermally shorts the blankets, reducing its effective emissivity. Two different sizes of I-pin tags were used: $3.5 \unit{mm}$ for normal tagging, and a longer $5 \unit{mm}$ tag when tagging through folds.

\subsubsection{Spectrograph Enclosure}
\label{sec:enclosure}
HPF will be placed in an enclosure in the Hobby-Eberly Telescope Spectrograph Room, which acts as a passive buffer to smooth out any short-time high-frequency temperature variations. This is a similar approach to the ESPRESSO spectrograph which uses two enclosures \citep{Pepe2014}. An enclosure was not used in the lab tests reported here.

During installation of the HPF enclosure at HET in November 2014, a temperature monitoring system was installed to monitor the performance of the enclosure and the HVAC system in the Spectrograph Room. Frequently updated plots of the temperatures inside and outside the enclosure can be found on the web\footnote{\url{http://gummiks.github.io/research/hpf_temps/}}. Initial results from this ongoing monitoring effort show that the enclosure effectively buffers out high-frequency ($<$$1 \unit{hour}$) temperature changes due to HVAC cycling in the Spectrograph Room. As expected, long-term trends (i.e. of order $\sim$$1 \unit{day}$ or longer) are observed to print through the enclosure: these slowly varying trends will be effectively corrected for by the HPF ECS.

\subsection{Vacuum Control}
\label{sec:vacuum}
Moderate to high vacuum is necessary for LN2-cooled astronomical instruments. HPF has a relatively large pumping volume of $\sim$$2.5 \unit{m^3}$, and a large surface area due to the multiple layers of MLI blanketing. Vacuum pressure is measured to the $<$$10^{-9} \unit{Torr}$ level using two independent sets of vacuum gauges: an MKS micropirani vacuum gauge and NexTorr pump (discussed further below). Initial vacuum pumping is achieved with a turbo-molecular vacuum pump mounted to an ISO-63 port on the vacuum cryostat, backed-up by an associated dry scroll pump. Long-term vacuum stability is maintained by two separate vacuum getter systems, working together to maintain ultimate vacuum pressures of $<$$10^{-7} \unit{Torr}$ long-term.

\begin{figure}
	\begin{center}
    \vspace{0.5cm}
		\includegraphics[width=0.98\columnwidth]{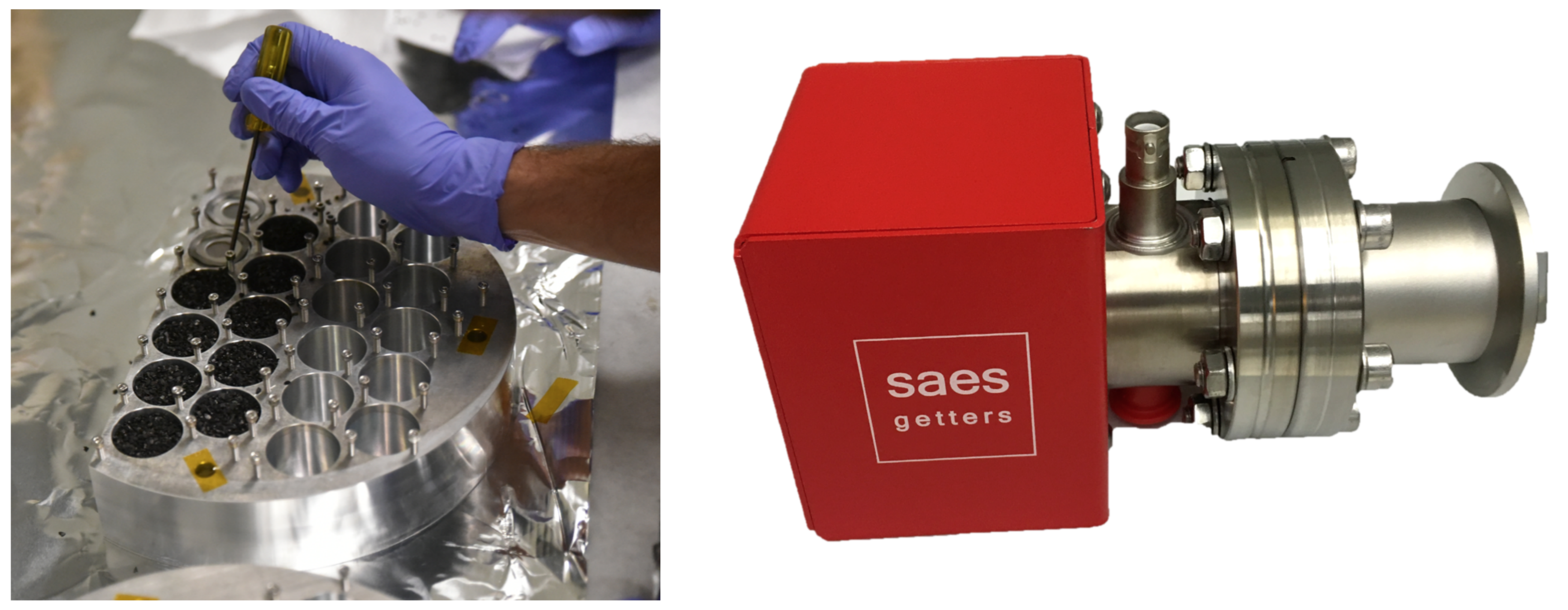}
	\end{center}
        \vspace{-0.2cm}
	\caption{Activated getters are used for long-term pressure stability: a) activated charcoal absorbers being populated in the charcoal getter-plate, which is directly mounted to the HPF LN2 tank; b) A NexTorr D-100 vacuum pump including an integrated hydrogen getter element and a small ion pump.}
	\label{fig:vacuum}
\end{figure}

The first getter system consists of activated charcoal absorbers similar to those used in the APOGEE cryostat \citep{Wilson2010}. The charcoal absorbers are kept in two custom-fabricated getter-plates (Figure \ref{fig:vacuum}a). The getter-plates are mounted to the ends of the LN2 tank, chilling the charcoal to LN2 temperatures, making for an effective cryopump. The getter-plates contain $\sim$$1 \unit{L}$ of activated charcoal each, which was sufficient for maintaining high quality vacuum in the APOGEE cryostat for several years \citep{Hearty2014}. An aluminum cap with an MLI blanket, covers the getters to radiatively isolate the getter-plates from the surroundings, while allowing ample room for molecules to reach the charcoal absorbers to be captured. The charcoal does not retain significant amounts of the condensed and physically adsorbed gases after the charcoal is warmed. Therefore, the turbo pump is used to pump out the desorbing gas during warm-up cycles, minimizing the gas load as the instrument is warmed and vented to atmosphere.

The second getter system is a commercial NexTorr D-100 pump from SAES getters. The pump combines a Non-Evaporative Getter (NEG) element with a small Sputter Ion Pump (SIP), making for a compact and light pump ($200 \times 75 \times 75\unit{mm}$) with a large pumping capacity, and high pumping speeds. The NEG element consists of a stack of porous sintered getter disks and has a large pumping capacity for active gases (hydrogen, water, carbon oxides, nitrogen etc.), while the SIP is efficient at removing inert gases not pumped by the NEG element (e.g., noble gases such as helium and argon). The SIP also offers a pressure reading, providing an independent measurement of the vacuum pressure within the system. The NEG removes gases at room temperature after a 1 hour activation period at $500 \unit{K}$ without any need for electric power to operate, making the pump robust to power outages.

\subsection{Temperature Monitoring and Control System}
\label{sec:tmc}
The HPF Temperature Monitoring and Control System (TMC) has been discussed in \cite{Hearty2014,Stefansson2016}, but we repeat salient points here for completeness.

\begin{figure}
	\begin{center}
    \vspace{0.5cm}
		\includegraphics[width=0.98\columnwidth]{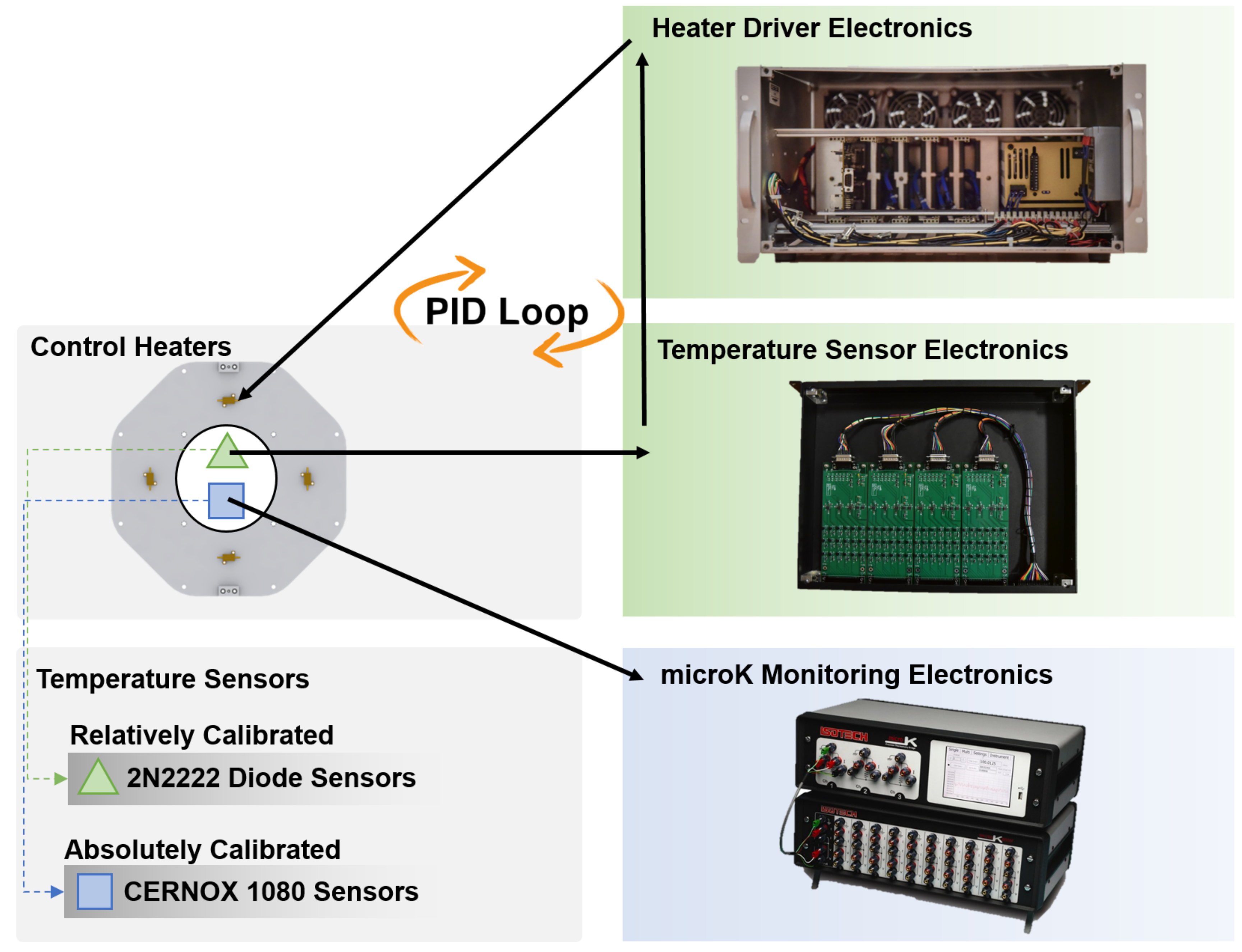}
	\end{center}
        \vspace{-0.2cm}
	\caption{Overview of the HPF Temperature Monitor and Control System (TMC). This system is responsible for actively controlling the temperature of the HPF radiation shield to the $180 \unit{K}$ set-point operating temperature.}
	\label{fig:tmc_panel}
\end{figure}

The TMC is responsible for keeping the radiation shield at the $180 \unit{K}$ operating temperature. Figure \ref{fig:tmc_panel} gives an overview of the system. It is composed of 14 resistive control heaters and corresponding temperature sensors, temperature sensing electronics, and temperature driving electronics. 

The TMC is composed of two completely separate temperature sensing systems: a dedicated Temperature Control System (TCS; green in Figure \ref{fig:tmc_panel}), and a dedicated Temperature Monitoring System (TMS; blue in Figure \ref{fig:tmc_panel}). Each of the 14 control heaters on the radiation shield have associated with them two temperature sensors, one from each temperature sensing system. The temperature measurements from the TCS sensors are used to drive the control heaters in a Proportional-Integral-Derivative (PID) feedback loop to a given set-point temperature. The temperature measurements from the TMS are used exclusively to monitor the performance of the PID loop and monitor the system for long-term drifts. The latter system has proven especially important for electronic diagnosing purposes. However, in the case of a TCS sensor failure, a TMS temperature sensor can be used to drive the control heater PID loop for a given control heater, allowing for a completely redundant system, making it robust to failure.

The TMS system uses a commercial MicroK 250 extremely low-noise thermometry bridge from Isotech. The TMS uses 20 CERNOX 1080 thermistor sensors for their temperature sensitivity at the $180 \unit{K}$ operating temperature of HPF. These sensors provide $0.1 \unit{mK}$ measurement precision at $180 \unit{K}$ when read out with the MicroK using an excitation current of $0.2 \unit{mA}$. The full set of 20 CERNOX temperature sensors were absolutely calibrated by Lake Shore Cryotronics. This system is used in a 4-wire configuration to minimize noise.

The TCS is a custom fabricated system. The temperature sensing relies on measuring the voltage drop of a diode under excitation by a low forward current ($\sim$$10 \unit{\upmu A}$), sensitive to temperature. The sensor used (2N2222) is a bipolar junction transistor, behaving like a diode when just two leads---the emitter and base pair---are excited. The 2N2222 temperature sensors are relatively calibrated to the absolutely calibrated TMS sensors.

\newpage

\section{Description of Stability Tests}
\label{sec:test_description}
Two long-time environmental stability tests were performed of the HPF cryostat and ECS at the Penn State Integration Lab: a cold test at the HPF $180 \unit{K}$ final operating temperature, and a warm test at an elevated operating temperature of nominally $300 \unit{K}$. The warm test was performed to empirically demonstrate the versatility and performance of the HPF ECS at warm temperatures, informing the ECS design and optimization for the NEID optical spectrograph, currently in development at Penn State. The cold test was performed in June 2016, and the warm test in Nov/Dec 2015. We note that the two tests have been discussed briefly in \cite{Stefansson2016} and \cite{Robertson2016}, but this manuscript contains a complete and detailed presentation of the results, with a full discussion of the ECS setup and design, including a SolidWorks 3D CAD model and a comprehensive parts list (Table \ref{tab:partslist}1), allowing for full reproducibility by the community.

Neither of the two tests used a passive thermal enclosure to buffer out the HVAC cycling in the lab environment. Therefore, these tests are more susceptible to short-term thermal fluctuations than during final science operations at the observatory. However, as is discussed in Section \ref{sec:Results}, the overall performance in a standard lab-condition environment is already excellent. 


The two tests are further summarized in Table~\ref{tab:test_compare}, and in the two subsections below. We first discuss the HPF cold test, performed after the warm test, as this test directly tested the functionality of the HPF ECS as designed.

\begin{table}
	\centering
	\caption{A comparison of the two long-term stability tests performed: the cold HPF-test, and the warm NEID concept-test. The final control setpoint is the temperature the control heaters were set to maintain on the radiation shield.  $T_{\textrm{failure}}$ is the equilibrium temperature the radiation shield would equalize to if the control heaters were turned off (e.g., during a power-outage or other failures).}
	\begin{tabular}{l c c}
	\hline\hline
	Stability test            & Cold (HPF)         & Warm (NEID)                           \\ \hline
	Nominal setpoint [K]      & $180$              & $300$                                 \\
    Final setpoint [K]        & $180.520$          & $302.500$                             \\
	$T_{\textrm{failure}}$ [K]& $\sim$$170$        & $T_{\textrm{Room}} \sim 293$          \\
    Ultimate heat sink        & LN2 tank           & Radiation to ambient                  \\
	Rad-shield thermal straps & Yes                & No                                    \\
	Thermal enclosure         & No                 & No                                    \\
    Control Sensors           & 2N2222             & CERNOX                                \\
    Dates performed  	      & June 2016          & Nov/Dec 2015                          \\
    \hline
	\end{tabular}
	\label{tab:test_compare}
\end{table}

\subsection{Long-term Temperature Stability at $180 \unit{K}$}
This stability test (see Table \ref{tab:test_compare}) was conducted at the HPF $180 \unit{K}$ operating temperature. The purpose of this test was to verify that HPF meets its short term temperature stability requirement of $3 \unit{mK}$ RMS per 24 hours, and its long-term stability requirement of $10 \unit{mK}$ RMS. This test was conducted in June 2016, with new and improved temperature control electronics from the warm-operation test. The control heaters were controlled with the TCS system, i.e. using the temperature readings from the 2N2222 sensors.

\subsection{Long-term Temperature Stability at $300 \unit{K}$}
This stability test (see Table \ref{tab:test_compare}) was conducted at $300 \unit{K}$ to assess the performance of the HPF ECS at warm temperatures, as a proof-of-concept study for the ECS for the NEID spectrograph. In this test all of the thermal straps to the radiation shield were disconnected from the LN2 tank. The control electronics were set to control the radiation shield at nominally $10 \unit{K}$ above ambient lab temperature, translating to a $\sim$$302.5 \unit{K}$ control setpoint, effectively converting the warm radiation shield to a distributed radiative heat sink. The LN2 tank was still filled, to ensure a level of high quality vacuum inside the spectrograph. This test was done in Winter 2015 (November-December). Due to problems with the TCS readout electronics, leading the 2N2222 sensors to not read out temperatures correctly, the control heaters were controlled with the TMS system.

\section{Stability Test Results}
\label{sec:Results}
\subsection{Temperature Stability at $180 \unit{K}$}
\begin{figure}
	\begin{center}
    \vspace{0.5cm}
		\includegraphics[width=0.98\columnwidth]{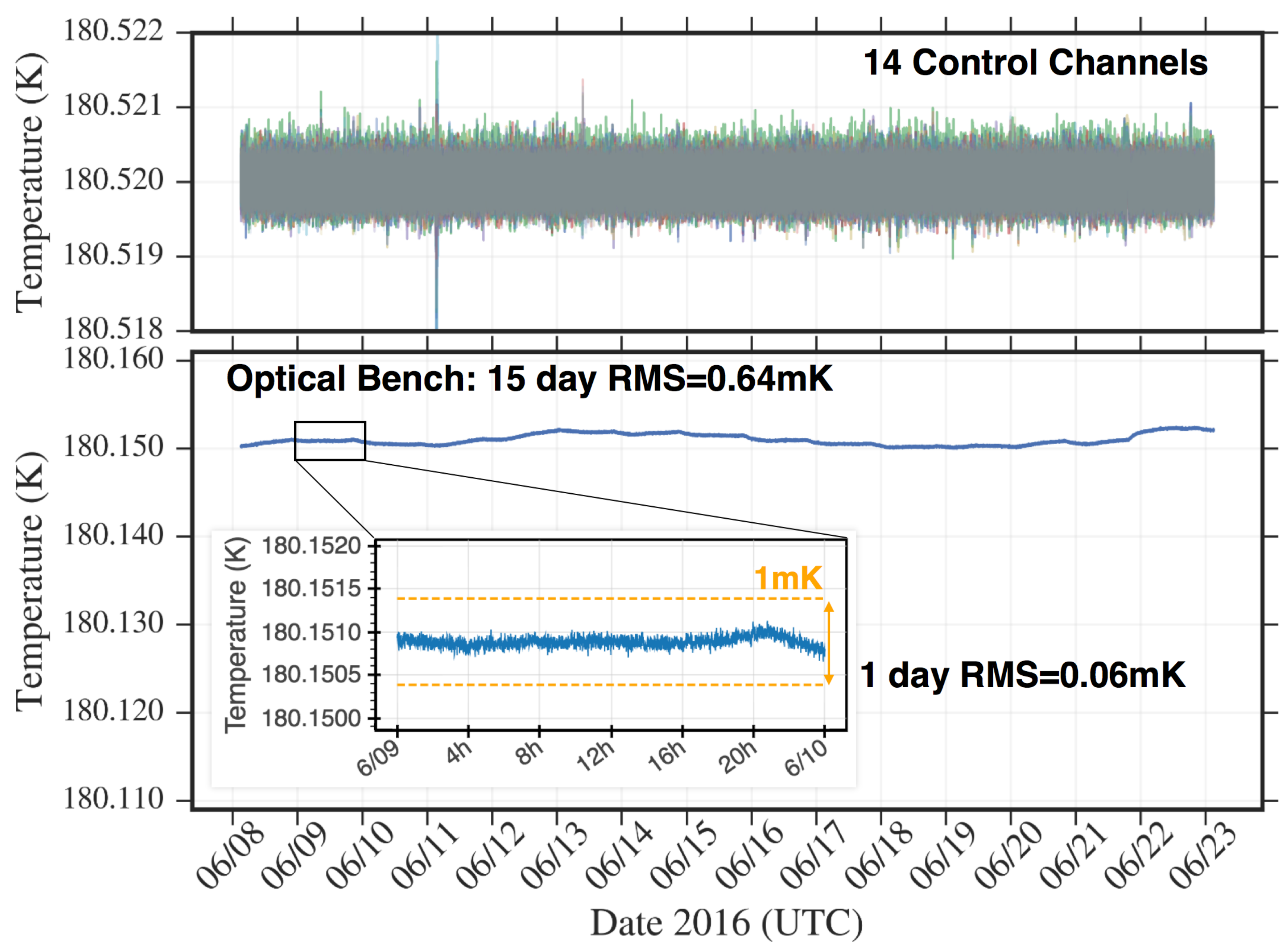}
	\end{center}
	\caption{Stability run at the $180 \unit{K}$ HPF operating temperature. The upper panel shows the temperatures as recorded by the 14 TCS control channels. The lower panel shows the temperature as recorded by the CERNOX sensor on the middle of the optical bench, with an RMS stability over 15 days of $\textrm{RMS}: 0.64 \unit{mK}$. The inset plot shows the temperature stability over 1 day, which is well within a $1 \unit{mK}$ band (dashed orange lines). The lower panel has been briefly described in \cite{Stefansson2016}.}
	\label{fig:hpf_cold_stability}
\end{figure}

Figure \ref{fig:hpf_cold_stability} shows temperatures of the control channels (under 2N2222/TCS control) in the upper panel, and the temperature as recorded by the TMS temperature sensor located in the center of optical bench (see location in Figure \ref{fig:ecs_overview}b) in the lower panel, over the 15-day testing period performed at the $180 \unit{K}$ HPF operating temperature. We see that throughout this period, the PID-loop of the TCS control channels maintains the temperature of the radiation shield within a $\pm 1 \unit{mK}$ control band. Throughout the 15-day testing period, the RMS temperature stability on the optical bench is $0.64 \unit{mK}$. The inset in Figure \ref{fig:hpf_cold_stability} shows that over a one day period the optical bench is stable to well within $1 \unit{mK}$. Table \ref{tab:bench} (2nd column) further summarizes the main temperature metrics from this stability test.

The small-scale long-term drift at the $\pm 1.3 \unit{mK}$ peak-to-valley level observed on the optical bench is due to a combination of two effects. 

The first effect is due to daily LN2-fill transients, which cause a brief temperature increase in the LN2 tank as pressurized warm LN2 from the storage Dewar is fed to the HPF LN2 tank. This results in a brief change in heat draw in the thermal straps, but as is shown in Figure \ref{fig:hpf_cold_stability}, the control heaters easily compensate for this transient on the radiation shield. We minimize this transient effect by depressurizing the LN2 tank by opening up to atmosphere for a few minutes both before and after each LN2 fill. Depressurizing the LN2 tank results in a lower LN2 saturation temperature, which we use to cancel out the otherwise observed increase in LN2 tank temperature. All of the daily LN2 fills during the 15-days in Figure \ref{fig:hpf_cold_stability} were done manually using this LN2 fill prescription. The typical drift observed by the bench due to this effect can be seen in the small $<$$1 \unit{mK}$ bump at end of 1-day inset in Figure \ref{fig:hpf_cold_stability}. This effect will be further minimized by installing a dedicated auto-fill system, currently being designed and fabricated.

The second effect is due to a slowly-varying drift in the absolute temperature calibration between the TMS and TCS temperature sensors, which leads to low-level ($\pm$$\leq$$0.5 \unit{mK}$) variability in temperatures as measured by the TMS sensors.  All temperature sensing pairs show the same drift, indicating the variability is not caused by the sensors themselves, as each sensor would diverge with a different amplitude and direction if it were inherently unstable.  Instead, we speculate the drift is due to the Seebeck effect \citep{seebeck1826}, wherein a temperature-dependent voltage arises at the interface of two metals in an electric circuit (e.g.~at the hermetic feed-through for the TMC electronics at the vacuum interface).  As a diagnostic for the performance of our TMC electronics, we installed a small number of extra channels in the TCS sensor suite where rather than installing a 2N2222 temperature sensor, we simply shorted the two cable leads together.  The voltages measured on these ``shorted" channels were closely correlated with the TMS/TCS calibration drifts.  Since these voltage changes were uncorrelated with the external laboratory temperatures--and since the laboratory temperature changes were too small to cause a significant resistance change in the TMC cables--we conclude that the Seebeck effect is the most likely explanation for the observed voltage and sensor calibration drifts.  If the Seebeck effect is in fact responsible for these drifts, it must originate in the TCS electronics, as the 4-wire temperature measurement scheme used by the TMS is designed specifically to avoid such effects.  While we demonstrate here that the current system performance exceeds our $1 \unit{mK}$ goal for HPF, we are currently working to better diagnose and mitigate the temperature changes caused by this phenomenon.

The operating temperature of HPF is nominally $180 \unit{K}$. However, as we see in Figure \ref{fig:hpf_cold_stability}, the actual temperature of the bench is around $\sim$$180.150 \unit{K}$, somewhat lower that the control channel temperatures controlling at $180.520 \unit{K}$. In the ideal case, where the optical bench is only radiatively coupled to the radiation shield and completely free of thermal conductive paths, the temperature of the bench and the radiation shield would be exactly the same. However, in practice this is hard to achieve, and the bench does have conductive heat transfer paths to the surroundings. These conductive paths include the G10 insulated hangers, and the forced cool-down lines. These two conductive paths feed together in roughly $3 \unit{W}$ of heat. The copper hanger straps have been sized to cancel out this heat input, but in practice, it is difficult to match the heat draw exactly, due to uncertainties in thermal resistances in the copper foils used.

We note that this test was not done using a passive temperature enclosure like the one HPF will use during science operations. This enclosure would improve the day-to-night temperature swings, but would not improve the stability results on a week-to-week basis. Therefore, during science operations for a given night, we expect to see better performance, than the 1-day RMSs reported here.


\subsection{Temperature Stability at $300 \unit{K}$}
\begin{figure}
	\begin{center}
    \vspace{0.5cm}
		\includegraphics[width=0.98\columnwidth]{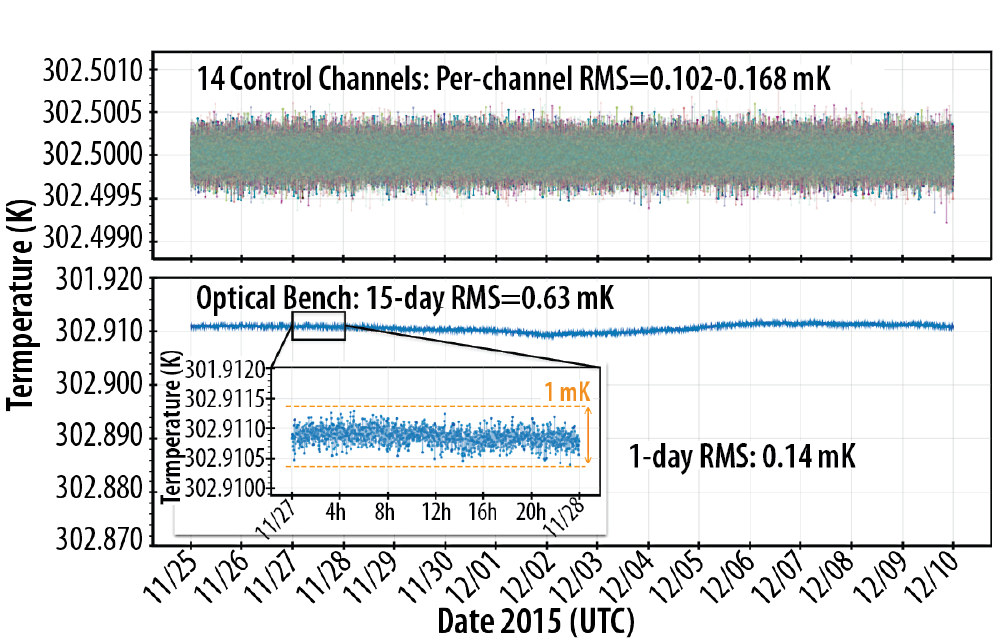}
	\end{center}
        \vspace{-0.2cm}
	\caption{The warm stability of HPF. The upper panel shows the temperature as recorded by the 14 control channels (under CERNOX/TMS control), set to maintain a temperature of $302.500 \unit{K}$. The lower plot shows the temperature of the optical bench over the 15-day period (15-day RMS: $0.63 \unit{mK}$) and an inset plot of 1-day temperature stability (1-day RMS: $0.14 \unit{mK}$). The lower panel has been briefly described in \cite{Robertson2016}.}
	\label{fig:hpf_warm_stability}
\end{figure}

Figure \ref{fig:hpf_warm_stability} shows temperatures of the control channels (under CERNOX/TMS control) in the upper panel, and the temperature of the optical bench in the lower panel over the 15-day testing period performed at nominally $300 \unit{K}$. Similar to Figure \ref{fig:hpf_cold_stability}, the RMS temperature stability is $\sim$$0.63 \unit{mK}$ over the full 15-day testing period. The short-term 1-day performance is even better, or at the $0.14 \unit{mK}$ level (see inset in Figure \ref{fig:hpf_cold_stability}). Similar to the cold test above, LN2 fills were performed daily. Table \ref{tab:bench} (3rd column) further summarizes the main temperature metrics from this stability test. 

\begin{table}
	\centering
	\caption{Comparison of optical bench stability from the cold ($T \sim 180 \unit{K}$), and warm ($T \sim 300 \unit{K}$) stability test. The 1-day range (peak-to-valley) and RMSs are calculated as the 1-day means from the 15 day datasets.}
	\begin{tabular}{l c c }
	\hline\hline
	  Stability test                & Cold (HPF) [mK] & Warm (NEID) [mK] \\
	 \hline
	 RMS: 1-day (mean)  & $ 0.17$          & $ 0.17$          \\
	 RMS: 15-day        & $ 0.64$          & $ 0.63$          \\
	 Range: 1-day (mean)& $ \pm 0.4$       & $ \pm 0.5$       \\
	 Range: 15-day      & $ \pm 1.3$       & $ \pm 1.5$       \\
	\hline
	\end{tabular}
	\label{tab:bench}
\end{table}

\subsection{Pressure Stability}
\begin{figure}
	\begin{center}
    \vspace{0.5cm}
		\includegraphics[width=1\columnwidth]{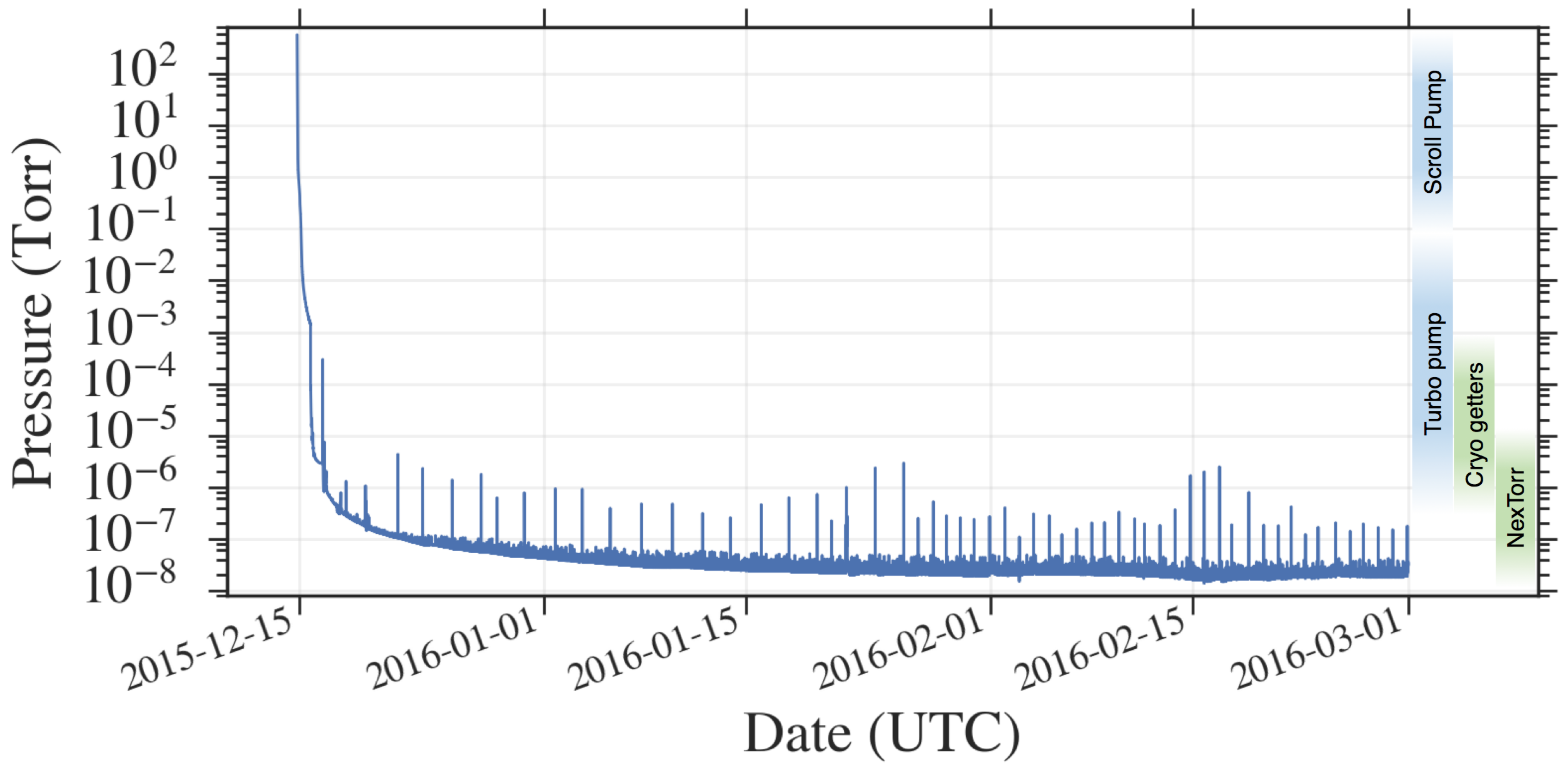}
	\end{center}
        \vspace{-0.2cm}
	\caption{HPF Long term pressure stability. The pressure regimes of each of the four HPF vacuum pumps are noted on the right side of the graph: the cryo getters and the NexTorr pump (green) are used during long-term operation, while the dry-scroll and turbo-molecular pumps (blue) are only used for initial vacuum pumping during maintenance. Briefly described in \cite{Stefansson2016}.}
	\label{fig:hpf_pressure}
\end{figure}

Figure \ref{fig:hpf_pressure} shows the long term pressure stability of HPF during a two and a half month long environmental stability run, from 15th of December 2015 to 1st of March, 2016, demonstrating better than  $10^{-7}\unit{Torr}$ pressure stability. This result indicates that the vacuum chamber well satisfies the $<$$10^{-6}\unit{Torr}$ pressure requirement for HPF.

Initial high pressures in Figure \ref{fig:hpf_pressure} correspond to the instrument being at ambient pressures. Initial rough pumping from ambient pressures to $\sim$$0.1 \unit{Torr}$ was achieved within a one day time frame using a large dry scroll pump, which was then switched over to a turbo-molecular pump to drop the pressure down to the $\sim$$10^{-6} \unit{Torr}$ level (Figure \ref{fig:hpf_pressure}). When the vacuum pressure reached $10^{-3} \unit{Torr}$, the LN2 tank was filled to start cryopumping with the activated charcoal getters. Ultimate pressures of $<$$10^{-7} \unit{Torr}$ were achieved by valving out the dry scroll and turbo-molecular pumps, and valving in the NexTorr pump, pumping out any residual gases. 

The periodic spikes in pressure seen in Figure \ref{fig:hpf_pressure} correspond to daily or bi-daily LN2 fills: as the warm nitrogen from the storage Dewar flows through the LN2 tank feed-lines, the cold feed-lines warm up and shake, causing a flux of molecules previously cryo-condensed to the feed-line surface to enter the vapor phase. This increases the vacuum pressure momentarily. However, the out-gassed particles are promptly pumped out by the charcoal getters and the NexTorr pump. These spikes will not affect the performance of HPF during science operations, as LN2 fills will be performed in the mornings, when HPF is not taking science data.

The HPF stainless steel cryostat has two main sources of vacuum gas-loads.

The first gas load is due to permeation and out-gassing from the Viton O-rings, which are often a major source of gas load, and can often limit the achievable base pressure of a vacuum system \citep{vacuumhandbook}. HPF has a total Viton O-ring length of $20.2 \unit{m}$ ($800 \unit{in}$), corresponding to a total gas load of $2\times 10^{-5} \unit{Torr~L^{-1} s^{-1}}$ due to O-ring permeation, assuming a standard Viton permeation rate of $2.5\times 10^{-8} \unit{Torr~L^{-1} s^{-1} in^{-1}}$ and a representative mixture of atmospheric gas at standard conditions. The effect on terminal vacuum pressure from this constant gas load is minimized by the high pumping power of the charcoal getters and the high pumping speed of the NexTorr pump.

\begin{figure}
	\begin{center}
    \vspace{0.5cm}
		\includegraphics[width=0.98\columnwidth]{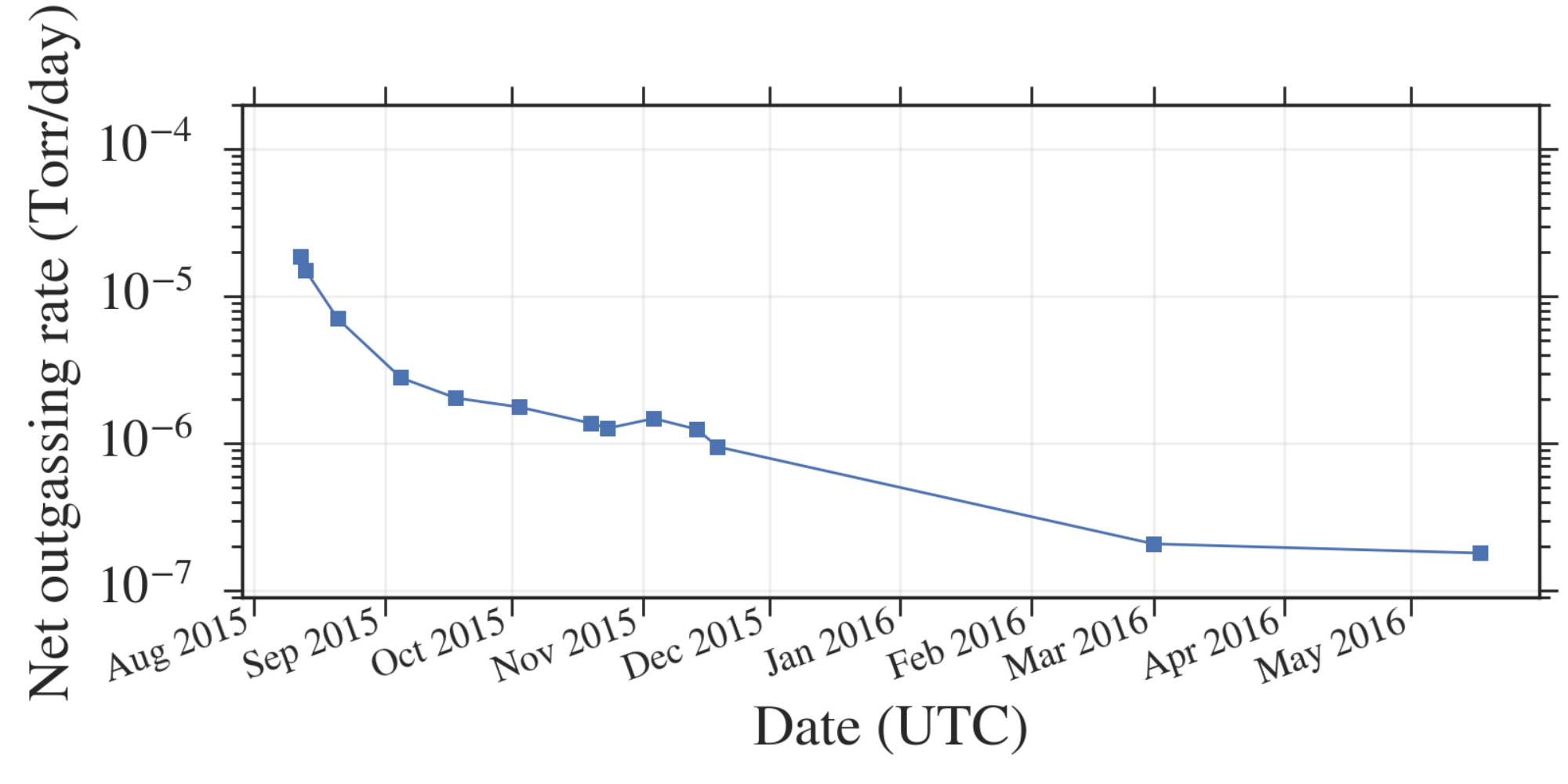}
	\end{center}
        \vspace{-0.2cm}
	\caption{Net outgassing rates as a function of time in the HPF cryostat, when only the charcoal getters are pumping. The high out-gassing rates in September 2015 are dominated by surface out-gassing of various molecules from the subsystems exposed to the vacuum within the cryostat (e.g., adhesives, desorbing of trapped water in MLI blankets, etc.). Hydrogen dominates after the September 2015 mark. In the recent months the out-gassing rate has plataued.}
	\label{fig:hpf_outgassing_rate}
\end{figure}

The second gas load is due to diffusion of hydrogen from the stainless steel cryostat walls. In our early stability experiments, this precluded vacuum stability at the $<$$10^{-7}\unit{Torr}$ level, as the out-gassing rate of hydrogen was on the order of $10^{-6}\unit{Torr~day^{-1}}$ or more (Figure \ref{fig:hpf_outgassing_rate}). Hydrogen is soluble in many materials, in particular in stainless steel, where it can get trapped in precipitates during fabrication, slowly out-gassing at room temperature \citep{vacuumhandbook}. Out-gassing of hydrogen from stainless steel slabs varies widely with surface treatments, and can be reduced by electro-polishing and oxidizing the the stainless steel exterior surface \citep{vacuumhandbook}. Additionally, baking the stainless steel walls at low temperatures of $150\unit{^{\circ}C}$ has been suggested to yield lower hydrogen out-gassing rates \citep{vacuumhandbook}. However, these options were not pursued for HPF as it was difficult to find an annealing chamber large enough to fit the HPF cryostat, and in addition these processes were incompatible with the Viton O-rings and sensitive electronic equipment already installed in the integrated cryostat.

Instead, the hydrogen out-gassing problem was solved by introducing a NexTorr pump with very high hydrogen pumping speeds of $100 \unit{L~s^{-1}}$, effectively pumping out the hydrogen instantaneously, not giving it time to build up pressure in the chamber. The out-gassing rate of hydrogen has decreased slowly with time, as is shown in Figure \ref{fig:hpf_outgassing_rate}, and in the recent months the rate has plateaued. The data presented in Figure \ref{fig:hpf_outgassing_rate} were taken by valving out the NexTorr, having only the charcoal getters pumping, allowing the hydrogen pressure to build overnight.

\newpage
\section{Discussion}
This paper has provided a description of an environmental control scheme capable of stabilizing a large vacuum cryostat for high-precision RV measurements at the sub-milli-Kelvin level over more than two weeks. The novel aspect of this control scheme is that the active temperature control elements are contained within the vacuum boundary, as opposed to outside. 

The system is robust against planned, and unplanned environmental transients. Planned transients include LN2 tank fills, introducing a mild pressure spike in the spectrograph vacuum. The performance of the system against unplanned environmental transients has been tested by monitoring the performance of the system in the current lab environment, which currently does not include a passive thermal enclosure, and is thus more susceptible to transient events. However, even without an enclosure we have demonstrated that this ECS is capable of sub-milli-Kelvin temperature stability long-term.

The failure modes of this ECS are designed to be benign. The full environmental control system is powered using an Uninterruptible Power Supply (UPS), allowing the system to stay powered for additional 6 hours in case of a full power outage. If the power outage is longer than 6 hours, the control channels will stop controlling at the given setpoint, and the system will naturally drift slowly to the natural back-to-nature failure temperature of nominally $\sim$$170 \unit{K}$. The vacuum will still be maintained by the passively pumping cryo-getters and NexTorr getters, for up to 4-5 days, limited by the passive hold-time of the LN2 tank.

The system is easily maintained, and serviced. The LN2 forced cool-down system, and warmup heater system allow for rapid cool-down and warm-up of the instrument, respectively. This saves time, and allows for quick warm-up and cool-down cycles. This is particularly important for NIR instruments like HPF, which tend to have long service times on the order of weeks.

The exquisite temperature stability of this ECS, along with the exquisite measurement precision of the temperature sensing electronics, has allowed us to identify further avenues of improvement. These error terms include the temperature change in the column pressure of LN2: if the liquid level of the LN2 in the LN2 tank is high, we see a slightly higher temperature in the LN2 tank due to column pressure effects. As LN2 boils off, we see the LN2 tank cool monotonically. This effect is small, and is minimized by reducing the changes in the LN2 liquid level to a minimum. Therefore, during science operations the LN2 tank will be filled each morning, to top off the tank with LN2. The boil-off rate of the LN2 itself can be reduced by improving the MLI blanketing by adding more layers, and improving how they cover up the complex edges of the LN2 tank where the getter-plates are located.

Another avenue for improvement we have identified is the conductive heat from the thin control heater and sensor wires. Even for $32 \unit{AWG}$ Phosphor Bronze low-thermal conductivity wire used for the temperature sensors, the conductive heat input is non-negligible---at the $\sim$$1 \unit{W}$ level for the total combined sensor and heater wiring. This heat is fed via conduction through the wire from room temperature to wire connector plates attached to the radiation shield. This effect has been mitigated by coupling one of the wire connector plates on the radiation shield to the LN2 tank via a thermal strap, sized to cancel out this input heat from the wires. Additionally, the stainless steel bellows hoses (see Figure \ref{fig:ecs_overview}) for the forced cool-down lines also act as a direct conductive path to the optical bench. The hoses put in power at the $\sim$$1 \unit{W}$ level only at one end of the optical bench. This has been addressed by adjusting the thickness of the hanger straps, to cancel out the input heat from the forced cool-down lines, minimizing thermal gradients on the bench.

This system is versatile and can be used to stabilize instruments at a range of temperatures, from $\sim$$77 \unit{K}$ to elevated room temperatures, with minimal modifications. The lower temperature bound is determined by the use of the LN2 heat-sink, while the upper bound is determined by the heating power of the control channel heaters, which in practice will be a few degrees above room temperature. We include a full SolidWorks assembly with this manuscript, along with a comprehensive parts list HPF ECS in Table \ref{tab:partslist}1 in the Appendix, to facilitate the adoption of this ECS in the broader astronomical community.

We are beyond the era where a single factor dominates the overall RV measurement precision. Now, comprehensive systems engineering error budgets are needed in order to have get an overview of where the error terms originate, and to allow us to identify avenues for improvement. For a long time instrument stability has been a tall pole in RV error budgets. The ideal environmental control scheme should be able to minimize that error term to a level comparable, or lower than other terms in the overall error budget. It is through the synergy between the lessons learned from past instruments and new technology development, that strides towards achieving $<$$ 1 \unit{mK}$ long-term spectrograph temperature stability are now being made. These advancements allow us to decouple obfuscating instrumental spectral shifts from the astrophysical, further illuminating the path to the 10~\cms~RV measurement precision necessary to detect Earth-twins.

\acknowledgments
We thank the anonymous referee for a thoughtful reading of the manuscript, and for useful suggestions and comments. This work was supported by funding from the Center for Exoplanets and Habitable Worlds. The Center for Exoplanets and Habitable Worlds is supported by the Pennsylvania State University, the Eberly College of Science, and the Pennsylvania Space Grant Consortium. GKS acknowledges support from the Leifur Eiriksson Foundation Scholarship. This work was supported by NASA Headquarters under the NASA Earth and Space Science Fellowship Program--Grant NNX16AO28H. This work was performed in part under contract with the California Institute of Technology (Caltech)/Jet Propulsion Laboratory (JPL) funded by NASA through the Sagan Fellowship Program executed by the NASA Exoplanet Science Institute. We acknowledge support from NSF grants AST1006676, AST 1126413, AST 1310885, and the NASA Astrobiology Institute (NNA09DA76A) in our pursuit of precision radial velocities in the NIR. We want to thank Zachary Prieskorn, Abe Falcone, and David Burrows at the Penn State Astronomy X-ray lab for loaning us their Residual Gas Analyzer to measure the gas constituents inside the HPF cryostat.

\appendix
\setcounter{table}{0}
\renewcommand{\thetable}{\Alph{section}\arabic{table}}
\section{Subsystem Equipment List}

\begin{deluxetable}{llllllp{4cm}}
\tabletypesize{\scriptsize}
\tablecaption{A list of major components constituting the environmental control system presented in this manuscript. Short comments are provided for each part/subsystem. More details can be found in the text, and through studying the SolidWorks 3D CAD model included with this manuscript.}

\tablehead{\colhead{Section} & \colhead{Subsystem} & \colhead{Component} & \colhead{Quantity} & \colhead{Vendor} & \colhead{Part Name} & \colhead{Comment}  \\ 
\colhead{} & \colhead{} & \colhead{} & \colhead{} & \colhead{} & \colhead{} & \colhead{} } 

\startdata
\ref{sec:cryostat}      & Cryostat                   & Cryostat shell           & 1                & PulseRay Inc.          & Custom             & Custom machining out of 304 stainless steel. Includes main frame, upper/lower lids, and vacuum penetrations. Includes trantorque shipping restraints for shipping.                                                            \\
                        & -                          & Vaccuum penetrations     & 5                & PulseRay Inc.          & Custom             & Custom stainless steel cutouts.                                                                                                                                                                                                  \\
                        & -                          & O-rings (Viton)          & 2                & PulseRay Inc.          & Custom length      & 2x O-rings for the upper and lower cryostat lids. Smaller pieces for vacuum penetrations. Viton for low outgassing.                                                                                                              \\
                        & -                          & Vibration isolation legs & 4                & Newport                & S-2000 Series      & 19.5in standard isolators with automatic leveling.                                                                                                                                                                               \\
                        & -                          & Shipping-pallet          & 1                & PulseRay Inc.          & Custom             & Custom design, with coil spring isolators and air skids.                                                                                                                                                                              \\
                        & -                          & Air skids                & 4                & Hovair Systems         & RSK16HD-4-MP6-MB42 & 4x airskids for shipping-pallet to facilitate shipping of the cryostat.                                                                                                                                                           \\
\ref{sec:radshield}     & Radiation shield           & Radiation shield shell   & 1                & PulseRay Inc.          & Custom             & Custom machining out of a combination of Aluminum alloy 6061-T6 and 3003. Includes main frame and upper/lower lids. 1/8" thick.                                                                                                \\
\ref{sec:opticalbench}  & Optical bench system       & Optical bench            & 1                & PulseRay Inc.          & Custom             & Monolithic piece of Kaiser select 6061-T6 aluminum. 3" thick, light weighted.                                                                                                                                                                              \\
                        & -                          & Optical bench hangers    & 4                & PulseRay Inc.          & Custom             & 4x stainless steel hanger mount assemblies to suspend the bench, and allow thermal contraction without inducing stress. G10 spacers are used to minimize thermal conduction.                                                                                                                                                                             \\
\ref{sec:thermalstraps} & Thermal Straps             & Radiation shield straps  & 16               & Custom/McMaster Carr   & Custom             & In house fabrication. Material: Multipurpose 110 Copper from McMaster Carr, 2" wide, 0.005" thick foils.                                                                                                                                       \\
                        & -                          & Hanger straps            & 4                & Custom/McMaster Carr   & Custom             & In house fabrication. Material: Multipurpose 110 Copper from McMaster Carr, 0.5" wide, 0.005" thick foils.                                                                                                                                     \\
\ref{sec:ln2tank}       & LN2 tank system            & LN2 tank                 & 1                & Precision Cryo Systems & Custom             & Custom machining out of Aluminum alloy 6061-T6, by Richard Gummer Precision Cryogenic Systems, Inc.                                                                                                                              \\
                        & -                          & LN2 level sensor         & 1                & American Magnetics     & Model 186          & Custom fabrication. 13" capacitance sensor.                                                                                                                                                                                   \\
\ref{sec:ln2tank}       & Pressure Regulator System  & Pressure Controller      & 1                & MKS Pressure Controller& 640B               &  Regulates the pressure in the LN2 tank, to stabilize the saturation temperature of LN2.                                                                                                                                                                                                             \\
\ref{sec:warmup}        & Warmup heater system       & Variacs                  & 2                & McMaster Carr          & 6994K24            & $120 \unit{V}$ input; $0-120\unit{V}$ output.                                                                                                                                                                                          \\
                        & -                          & $50 \unit{\Omega}$ heaters  & 16            & Vishay Dale            & RH05050R00FC02     & Heaters rated at $50 \unit{\Omega}$, $50 \unit{W}$ each.                                                                                                                                                                        \\
                        & -                          & Warmup heater wire       &                  & Mouser Electronics     & Alpha Wire 5853    & Heavy duty stranded wire $26 \unit{AWG}$, PTFE coated for low outgassing.                                                                                                                                                                    \\
\ref{sec:mli}           & MLI blankets               & Aluminized Mylar         &                  & Multek                 & 146466-003         & $0.25 \unit{Mil}$ double sided aluminized Mylar, 48" wide. Insulating material for MLI blankets.                                                                                                                                              \\
                        & -                          & White tulle              &                  & Tulle Source           & White Tulle        & White Nylon Tulle Fabric, 54" width. Used as spacer material for MLI blankets.                                                                                                                                                   \\
                        & -                          & Nylon tag gun            & 2                & Jay Industrial Corp    & Tagstar XB         & Used to hold MLI blankets together. Tags: $3.5 \unit{mm}$, and $5 \unit{mm}$.                                                                                                                                                    \\
                        & -                          & Aluminized Tape          &                  & ULine                  & S-15881SIL         & Silver aluminized Mylar tape. For added strength at tag locations.                                                                                                                                                               \\
\ref{sec:enclosure}     & Thermal Enclosure          & Spectrograph enclosure   & 1                & Bally Enclosures       & Custom             & Custom size (16'-10"x14'-11"x10'-3" exterior dimensions), used to passively buffer high frequency temperature fluctuations. Without insulated floor for added temperature buffering.\\
\ref{sec:vacuum}        & Vaccuum Control System     & Activated charcoal       & 2L               & Fischer Scientific     & 05-685B            & Activated Carbon 6-14 mesh, lot-no: PF131AE.                                                                                                                                                                                     \\
                        & -                          & Getter holders           & 2                & PulseRay Inc.          & Custom             & In house fabrication. Two in total, one for each end of LN2 tank.                                                                                                                                                                 \\
                        & -                          & NexTorr D-100            & 1                & SAES Getters           & D-100              & An integrated NEG hydrogen getter element and a small ion pump. Low-power, and long-term stable vacuum pump. \\ 
                        & -                          & Turbo pump               & 1                & Agilent                & 9698901            & Turbo pump (V 81M ISO 63), used to get to $\sim$$10^{-6} \unit{Torr}$ range in pressure.                                                                                                                                          \\
                        & -                          & Tri Scroll pump          & 1                & Agilent                & PTS03001UNIV       & Tri-Scroll 300 I Phase, used as rough vacuum pump and backing pump for turbo pump.                                                                                                                                               \\
                        & -                          & Gate valve               & 1                & Agilent                & X3202-60004        & Aluminum ISO-63 gate valve for turbo pump.                                                                                                                                                                                       \\
                        & -                          & Vaccuum gauge            & 1                & MKS                    & 972B-21034         & Micropirani vacuum gauge.                                                                                                                                                                                                        \\
\ref{sec:tmc}           & Temperature Control System & Heater panel baseplates  & 14               & Custom/McMaster Carr   & Custom/9246K13     & In house fabrication out of Aluminum alloy 6061-T6.                                                                                                                                                                              \\
                        & -                          & Control heaters          & 56               & Vishay Dale            & RH005              & 4 heaters per control channel, rated at $150 \unit{\Omega}$, $5\unit{W}$ each.                                                                                                                                                   \\
                        & -                          & Control heater wire      &                  & LakeShore              & LakeShore WHD-30   & Heavy duty stranded wire $30 \unit{AWG}$.                                                                                                                                                                                          \\
                        & -                          & Thermometry Bridge       & 1                & Isotech                & microK-250         & MicroK 250 Precision Thermometry Bridge. Used as a dedicated monitoring channel.                                                                                                                            \\
                        & -                          & 10 channel scanner       & 2                & Isotech                & microsKanner       & 2x 10 channel scanners for MicroK, used with the CERNOX sensors.                                                                                                                                                                            \\
                        & -                          & Reference resistor       & 1                & Isotech                & Model 456          & Model 456 Temperature Controlled Fixed Resistor ($400 \unit{\Omega}$).                                                                                                                                                           \\
                        & -                          & CERNOX sensors           & 20               & LakeShore Cryotronics  & CERNOX 1080        & Absolutely calibrated by LakeShore.                                                                                                                                                                                              \\
                        & -                          & Cryogenic wire           &                  & LakeShore Cryotronics  & LakeShore WDT-32   & Dual twist 32 AWG wire, cryogenic and vacuum compatible.                                                                                                                                                                                \\
                        & -                          & Vacuum hermetic, 55 pin  & 4                & Sierra IC Inc.         & PT07H-22-55P       & 55 pin vacuum hermetic for control heater and sensor electronics.                                                                                                                                                                               \\
                        & -                          & Vacuum hermetic, 32 pin  & 3                & Sierra IC Inc.         & PT07H-18-32P       & 32 pin vacuum hermetic for control heater and sensor electronics.                                                                                                                                                                               \\
                        & -                          & 2N2222 sensors           & 35               & ON Semiconductor       & JANTXV2N2222A      & Bipolar junction transistor used in a diode configuration for precise temperature sensing.                                                                                                                                                                                                                               \\
                        & -                          & Thermal epoxy            &                  & EpoTek                 & EpoTek 7110        & Thermally conductive, but electrically insulating epoxy for 2N2222 sensors.                                                                                                                                                                                    \\
                        & -                          & Custom TMC Electronics   & 1                & Custom                 & Custom             & Detailed design and description will be provided in Robertson et al.~(2017).                                                                                                                                                              \\
\enddata
\label{tab:partslist}
\end{deluxetable}

\bibliographystyle{yahapj}
\bibliography{references}

\end{document}